\documentclass[superscriptaddress,twocolumn,amssymb,nofootinbib,10pt]{revtex4-2}
\usepackage{amsmath, amsthm, amssymb} 

\usepackage{graphicx}
\usepackage{color}\usepackage{amsmath}
\usepackage[compat=1.1.0]{tikz-feynman}
\usepackage{tikz}
\usetikzlibrary{arrows,shapes}
\usetikzlibrary{trees}
\usetikzlibrary{matrix,arrows} 			
\usetikzlibrary{positioning}				
\usetikzlibrary{calc,through}				
\usepackage{pgffor}							

\newcommand{\beq}{\begin{equation}}
\newcommand{\eeq}{\end{equation}}
\newcommand{\bqa}{\begin{eqnarray}}
\newcommand{\eqa}{\end{eqnarray}}

\def\sumint{\hbox{$\sum$}\!\!\!\!\!\!\int}
\def\square{\vcenter{\vbox{\hrule height.4pt
          \hbox{\vrule width.4pt height4pt
          \kern4pt\vrule width.3pt}\hrule height.4pt}}}

\begin{document}

\title{Thermodynamics and quark condensates of three-flavor QCD at low temperature}

\author{Jens O. Andersen}
\email{andersen@tf.phys.ntnu.no}
\affiliation{Department of Physics, Faculty of Natural Sciences,NTNU, 
Norwegian University of Science and Technology, H{\o}gskoleringen 5,
N-7491 Trondheim, Norway}

\author{Qing Yu}
\email{yuq@cqu.edu.cn}
\affiliation{Department of Physics, Chongqing University, Chongqing 401331, People’s Republic of China}
\affiliation{Department of Physics, Faculty of Natural Sciences,NTNU, 
Norwegian University of Science and Technology, H{\o}gskoleringen 5,
N-7491 Trondheim, Norway}
\author{Hua Zhou}
\email{zhouhua@cqu.edu.cn}
\affiliation{Department of Physics, Chongqing University, Chongqing 401331, People’s Republic of China}
\affiliation{Department of Physics, Faculty of Natural Sciences,NTNU, 
Norwegian University of Science and Technology, H{\o}gskoleringen 5,
N-7491 Trondheim, Norway}
\date{\today}

\begin{abstract}
We use three-flavor chiral perturbation theory ($\chi$PT) to calculate the 
pressure, light and $s$-quark condensates of QCD in the 
confined phase at finite temperature to ${\cal O}(p^6)$ in the low-energy expansion.
We also include electromagnetic effects to order $e^2$, where the electromagnetic coupling
$e$ counts as order $p$.
Our results for the pressure and the condensates
suggest that $\chi$PT converges very well for temperatures 
up to approximately 150 MeV.
We combine $\chi$PT and the Hadron Resonance Gas (HRG) model by adding 
heavier baryons and mesons. Our results are compared with lattice simulations and 
the agreement is very good for temperatures below {170} MeV, 
{in contrast to the results from $\chi$PT which agree with the lattice only up to
$T\approx120$ MeV.}
Our value for the chiral crossover temperature is  {160.1 MeV}, which compares favorably to
the lattice result of $157.3$ MeV.

\end{abstract}
\keywords{Dense QCD,
chiral transition, }

\maketitle

\section{Introduction}
In massless QCD with three flavors, the QCD Lagrangian has a global $SU(3)_L\times SU(3)_R\times U(1)_B$ symmetry in addition to a local $SU(N_c)$ gauge symmetry.
In the vacuum, this symmetry is broken down to $SU(3)_V\times U(1)_B$ via the formation
of a quark condensate, which gives rise to eight massless Goldstone bosons, 
the charged and neutral pions, the charged and neutral kaons, and the eta.
In nature, this symmetry is explicitly broken by finite quark masses down to
$SU(2)_V\times U(1)_Y\times U(1)_B$ giving rise to pseudo-Goldstone bosons whose masses
are small compared to the typical hadronic scale.
The low-energy effective theory that describes the pseudo-Goldstone bosons is chiral
perturbation theory ($\chi$PT), which is based only on the global symmetries of QCD and the low-energy degrees of freedom~\cite{wein,gasser1,gasser2}. 
It therefore provides a model-independent framework for describing the low-energy dynamics of QCD. 

The original formulation of $\chi$PT was in the strong sector. Gasser and Leytwyler 
developed a consistent power counting scheme such that the effective Lagrangian can be written as
an infinite series of terms in a low-energy expansion. The leading-order Lagrangian
is simply the nonlinear sigma model. The next-to-leading order Lagrangian for two flavors
was derived in Ref.~\cite{gasser1} and for three flavors in Ref.~\cite{gasser2}.
At next-to-next-to leading order, the effective Lagrangian was derived in
Refs.~\cite{fearing,bein,bein2}. A review of the phenomenology of chiral perturbation theory was
given in Ref.~\cite{bijnensreview}.

In the strong sector, the charged and neutral pions have the same tree-level masses.
A mass difference between the $u$ and the $d$ quarks, produces
isospin breaking effects in hadron masses. For pions, this effect is second order
in $m_u-m_d$. For charged and neutral kaons it turns out that their mass difference 
is linear in the quark mass difference $m_u-m_d$. However, there is another important source of the
mass differences between the neutral and charged mesons, namely the effects of virtual photons.
The leading electromagnetic effects of order $e^2$ were first included in Ref.~\cite{e2chi}, while
the systematic inclusion of the effects of
virtual photons in $\chi$PT at next-to-leading order, i.e. ${\cal O}(e^2p^2)$ and
${\cal O}(e^4)$ was carried out in Refs.~\cite{urech1,meis1,meis2,urech2}.
The power counting rule in $\chi$PT developed in~\cite{urech1} is such that $e$ counts as order $p$. 

Given the success of chiral perturbation theory at $T=0$, one may hope that it also provides
a good description of the QCD thermodynamics at low temperature. In the chiral limit, the
pions are massless and their typical momenta are of order $T$. If $T$ is sufficiently small,
the low-energy expansion ought to converge. Again, in the massless limit, the pion decay constant
$f$ is the only scale that appears in the leading-order Lagrangian. Up to corrections
given by the low-energy constants that appear at higher orders in the low-energy expansion,
$T/f$ is the expansion parameter of $\chi$PT. In a series of papers, the
low-temperature expansions of the pressure ${\cal P}$ and quark condensate $\langle\bar{q}q\rangle$
in two-flavor
$\chi$PT were calculated and show the expected form~\cite{gasser3,gas4,finitet}
\bqa
\label{p8}
{\cal P}
&=&{\pi^2T^4\over30}\left[1+{1\over36}{T^4\over f^4}
\log{\Lambda_p\over T}
+{\cal O}\left({T^6\over f^6}\right)
\right],
\\ \nonumber
\langle\bar{q}q\rangle&=&\langle\bar{q}q\rangle_0
\left[1-{1\over8}{T^2\over f^2}
-{1\over384}{T^4\over f^4}-{1\over288}{T^6\over f^6}
\log{\Lambda_q\over T}
\right.\\ &&\left.
+{\cal O}\left({T^8\over f^8}\right)
\right]\;,
\eqa
where $\langle\bar{q}q\rangle_0$ is condensate in the vacuum,
$\Lambda_p=275\pm 65$ MeV and $\Lambda_q=470\pm110$ MeV~\cite{finitet}.
$\Lambda_p$ and $\Lambda_q$ depend on the low-energy constants $\bar{l_i}$,
which up to a factor are the running couplings $l_i^r(\Lambda)$ evaluated at the scale 
$\Lambda=m$, $m$ being the (bare) pion mass. 
The expansions show good convergence properties for temperatures up to approximately 
140 MeV~\cite{finitet}.

However, at high enough temperature, $\chi$PT ceases to be valid since 
other degrees of freedom are excited and one must use other methods. 
Lattice Monte Carlo techniques is a first-principles method that can be used to study 
finite-temperature QCD:
At zero (and small) baryon chemical potential, one can carry out lattice simulations to
calculate thermodynamic quantities such as the pressure and interaction measure as well
as the approximate
order parameters that characterize confinement and chiral symmetry breaking, namely
the Polyakov loop and the quark condensates. For physical quark masses 
and two quark flavors, the transition is a smooth crossover at a transition temperature of around 155 MeV~\cite{tc1,tc2,tc3,tc4}.

The Hadron resonance gas (HRG) model treats finite-temperature QCD as a gas
of non-interacting hadrons and their resonances. As $T$ gets higher, it is necessary
to include more and more particles, and typically one has included
the approximately 200 hadrons below 2.5 GeV.
It can be easily generalized to finite chemical potentials as well ``distorted"
by using not the physical masses but masses that take into account lattice discretization effects.
It has also been combined with results from two-flavor $\chi$PT
by adding the contributions from heavier hadrons~\cite{tc3}.  
Comparing predictions for e.g. the pressure and the quark condensate of lattice QCD
and the HRG model, one finds, perhaps surprisingly,
very good agreement given the fact that the latter 
does not include interactions 
(unless combined with e.g. $\chi$PT)~\cite{tc3,hrg1,hrg2,bla,peter}.

Finite temperature calculations within $\chi$PT including electromagnetic effects are
scarce. In Ref.~\cite{nicola}, the authors calculate the quark condensates at NLO
in two-flavor and three-flavor $\chi$PT.
In Ref.~\cite{massnicola}, they calculate the pole masses
and the damping rate for the charged pion in two-flavor $\chi$PT at LO in the classes of
covariant and Coulomb gauges. While the pole mass is gauge-fixing independent in the
two classes of gauges and coincide, the damping rate depends on the gauge.
In particular, the damping rate in covariant gauge is negative indicating an instability.
This is reminiscent of the old problem of the gauge dependence of the
gluon damping rate in hot QCD. The problem was solved by Braaten and Pisarski
who realized that a one-loop calculation is incomplete and that one must use
effective propagators and vertices to obtain a complete leading-order result~\cite{bpdamp,softampl}. This is summarized in a non-local effective
Lagrangian that upon expansion generates the correction terms~\cite{gil1,gil2}.
This Lagrangian has been generalized to all temperatures and densities in Ref.~\cite{ericallt}
and can possibly be used to resolve the gauge dependence 
of the damping rate in $\chi$PT.

In the present paper, we consider three-flavor $\chi$PT at finite temperature including
electromagnetic effects to leading order in $e^2$. We calculate the pressure and the
quark condensates to ${\cal O}(p^6)$. In order to extend the validity of our calculations
to higher temperatures, we combine the results from $\chi$PT and the hadron resonance gas model.
The latter has enjoyed considerable success in describing the thermodynamics of 
low-temperature QCD as obtained from the lattice.
The paper is organized as follows. In Sec.~II, we briefly discuss the chiral Lagrangian. In. Sec.~III, we calculate the pressure to ${\cal O}(p^6)$ in the low-energy expansion. In Sec.~IV, we 
discuss the extension of the chiral Lagrangian to include the effects of electromagnetic
interactions. In Sec.~V, the pressure is again calculated to ${\cal O}(p^6)$ in the low-energy expansion. In Sec.~VI, we calculate the quark condensates
while in Sec.~VII we briefly discuss the hadron resonance gas model.
In Sec.~VIII, we present and discuss our numerical results. 
We have included four appendices providing the reader with definitions and
useful calculational details. In particular, we calculate the quark condensate at $T=0$
including electromagnetic effects, which is required in the calculation of the
finite-temperature dependent quark condensates.

\section{Chiral Lagrangian}
In massless three-flavor QCD, the Lagrangian has a global $SU(3)_L\times SU(3)_R$
symmetry in addition to the global $U(1)_B$ baryon symmetry and the local $SU(N_c)$
gauge symmetry. In the vacuum, this symmetry is broken to 
$SU(3)_V$ by the formation of a quark condensates.
For two massless and one massive quark, the symmetry is 
$SU(2)_L\times SU(2)_R$, which is broken to $SU(2)_V$ in the vacuum.
For two degenerate light quarks and one massive quark, this symmetry is explicit broken
to $SU(2)_V$.
If the two quarks are nondegenerate, we have three $U(1)$ symmetries, one for each quark 
flavor.

Chiral perturbation theory is a low-energy effective theory of QCD which is based on the global
symmetries and relevant degrees of freedom~\cite{wein,gasser1,gasser2}. 
For three-flavor QCD, the degrees of freedom are the eight mesons:
three pions, four kaons, and the $\eta$. In the chiral Lagrangian each factor of a quark mass
counts two powers of momentum
and each factor of a derivative counts one power of momentum.
The leading-order Lagrangian is given by~\cite{gasser2}
\bqa
{\cal L}_2&=&
{1\over4}f^2\langle\partial_{\mu}\Sigma\partial^{\mu}\Sigma^{\dagger}\rangle
+{1\over4}f^2\langle\chi^{\dagger}\Sigma+\Sigma^{\dagger}\chi\rangle
\;,
\eqa
where $\langle A\rangle$ denotes the trace of a matrix $A$ {in flavor space}, $f$ is the bare pion decay 
constant, and $\chi$ is given in terms of the quark mass matrix
\bqa
\chi&=&2B_0\,{\rm diag}(m_u,m_d,m_s)\;.
\eqa
Finally, 
\bqa
\Sigma &=& \exp\left[i{\lambda_a\phi_a\over f}\right]\;,
\eqa
with $\phi_a$ being the meson fields parameterizing the Goldstone manifold and
where $\lambda_a$ are the Gell-Mann matrices that satisfy $\langle\lambda_a\lambda_b\rangle=2\delta_{ab}$.

Expanding the Lagrangian ${\cal L}_2$  to second order in the fields 
$\phi_a$, we find
\bqa
\nonumber
{\cal L}_2^{\rm quadratic}&=&
\partial_{\mu}\pi^+\partial^{\mu}\pi^- -m_{\pi,0}^2\pi^+\pi^-
\\ && \nonumber
+{1\over2}\partial_{\mu}\pi^0\partial^{\mu}\pi^0
-{1\over2}m_{\pi,0}^2(\pi^0)^2
\\ && \nonumber
+\partial_{\mu}K^+\partial^{\mu}K^- -m_{K^{\pm},0}^2K^+K^-
\\  \nonumber && 
+\partial_{\mu}K^0\partial^{\mu}\bar{K}^0 -m_{K^0,0}^2K^0\bar{K}^0
\\ && 
+{1\over2}\partial_{\mu}\eta\partial^{\mu}\eta
-{1\over2}m_{\eta,0}^2\eta^2\;,
\eqa
where the meson fields are expressed in terms of $\phi_a$ as 
\bqa
\pi^{\pm}&=&{1\over\sqrt{2}}(\phi_1\mp i\phi_2)\;,\\
\pi^0&=&\phi_3\;,\\
K^{\pm}&=&{1\over\sqrt{2}}(\phi_4\mp i\phi_5)\;,\\
K^{0}/\bar{K}^0&=&{1\over\sqrt{2}}(\phi_6\mp i\phi_7)\;,\\
\eta&=&\phi_8\;.
\eqa
The tree-level masses are
\bqa
m_{\pi,0}^2&=&B_0(m_u+m_d)\;,\\
m_{K^{\pm},0}^2&=&B_0(m_u+m_s)\;,\\
m_{K^0,0}^2&=&B_0(m_d+m_s)\;,\\
m_{\eta,0}^2&=&{B_0(m_u+m_d+4m_s)\over3}\;.
\eqa
Since we are working in the isospin limit, there is no mixing between $\pi^0$ and $\eta$.
As long as $e=0$, the charged and neutral kaons have the same bare mass
which is denoted by $m_{K,0}$.
\begin{widetext}
The quartic terms of the Lagrangian ${\cal L}_2$ contains a large number of terms.
They can conveniently be written as 
\bqa\nonumber
{\cal L}_2^{\rm quartic}
&=&{m_{\pi,0}^2\over24f^2}(\pi^0)^4
+{m_{\pi,0}^2\over12f^2}(\pi^0)^2\eta^2
+{m_{\pi,0}^2\over6f^2}\pi^+\pi^-\eta^2
+{1\over216f^2}(16m_{K,0}^2
-7m_{\pi,0}^2)\eta^4\\ && \nonumber
-{1\over6f^2}\left[2(\pi^0)^2\partial_{\mu}\pi^+\partial^{\mu}\pi^-
+2\pi^+\pi^-\partial_{\mu}\pi^0\partial^{\mu}\pi^0
-m^2_{\pi,0}\pi^+\pi^-(\pi^0)^2\right]\\ && \nonumber
-{1\over6f^2}\pi^+\pi^-\left[2\partial_{\mu}\pi^+\partial^{\mu}\pi^--m_{\pi,0}^2\pi^+\pi^-
\right]-{1\over6f^2}K^+K^-\left[2\partial_{\mu}K^+\partial^{\mu}K^-
-m_{K,0}^2K^+K^-\right]\\ && \nonumber
-{1\over6f^2}\pi^+\pi^-\left[\partial_{\mu}K^+\partial^{\mu}K^--m_{K,0}^2K^+K^-\right]
-{1\over6f^2}K^+K^-\left[\partial_{\mu}\pi^+\partial^{\mu}\pi^-
-m_{\pi,0}^2\pi^+\pi^-\right]\\ && \nonumber-{1\over6f^2}
\left[2K^0\bar{K}^0\partial_{\mu}K^0\partial^{\mu}\bar{K}^0-m_{K,0}^2(K^0\bar{K}^0)^2\right]\\ && \nonumber
-{1\over12f^2}\left[
K_0\bar{K}_0\partial_{\mu}\pi_0\partial^{\mu}\pi_0
+{(\pi^0)}^2 \partial_{\mu}K_0\partial^{\mu}\bar{K}_0-
(m_{\pi,0}^2+m_{K,0}^2){(\pi^0)}^2 K_0\bar{K}^0\right]\\ && \nonumber
-{1\over12f^2}\left[K^+K^-\partial_{\mu}\pi_0\partial^{\mu}\pi_0
+{(\pi^0)}^2 \partial_{\mu}K^+\partial^{\mu}K^-
-(m_{\pi,0}^2+m_{K,0}^2)
{(\pi^0)}^2 K^+K^-\right]\\ && \nonumber
-{1\over6f^2}\left[K^0\bar{K}^0\partial_{\mu}\pi^+\partial^{\mu}\pi^-
+\pi^+\pi^-\partial_{\mu}K^0\partial^{\mu}\bar{K}^0
-(m_{\pi,0}^2+m_{K,0}^2)\pi^+\pi^-K^0\bar{K}^0\right]\\ && \nonumber
-{1\over6f^2}\left[K^+K^-\partial_{\mu}K^0\partial^{\mu}\bar{K}^0
+K^0\bar{K}^0\partial_{\mu}K^+\partial^{\mu}K^-
-2m_{K,0}^2K^+K^-K^0\bar{K}^0\right]\\ && \nonumber
-{1\over12f^2}\left[3K^+K^-\partial_{\mu}\eta\partial^{\mu}\eta
+3\eta^2\partial_{\mu}K^+\partial^{\mu}K^-
+(m_{\pi,0}^2-3m_{K,0}^2)K^+K^-\eta^2\right]\\ &&
-{1\over12f^2}\left[
3K^0\bar{K}^0\partial_{\mu}\eta\partial^{\mu}\eta
+3\eta^2\partial_{\mu}K^0\partial^{\mu}\bar{K}^0
+(m_{\pi,0}^2-3m_{K,0}^2)K^0\bar{K}^0\eta^2\right]\;,
\eqa
\end{widetext}
where we have omitted terms that do not
contribute to the pressure or quark condensates
at two loops in the isospin limit.

At next-to-leading order in the low-energy expansion, 
there are 12 terms in the chiral Lagrangian~\cite{gasser2}. The terms that are
relevant for the present calculations are
\bqa
\nonumber
{\cal L}_4&=&
L_4\langle\partial_{\mu}\Sigma^{\dagger}\partial^{\mu}\Sigma\rangle
\langle\chi^{\dagger}\Sigma+\chi\Sigma^{\dagger}\rangle
\\ && \nonumber
+L_5\langle\left(\partial_{\mu}\Sigma^{\dagger}\partial^{\mu}\Sigma\right)
\left(\chi^{\dagger}\Sigma+\chi\Sigma^{\dagger}\right)\rangle
\\ \nonumber
&&
+L_6\langle\chi^{\dagger}\Sigma
    +\chi\Sigma^{\dagger}\rangle^2
    +L_{7}\langle\chi\Sigma^{\dagger}-\chi^{\dagger}\Sigma\rangle^{2}
\\ && 
+L_8\langle\chi^{\dagger}\Sigma \chi^{\dagger}\Sigma
+  \chi\Sigma^{\dagger}\chi\Sigma^{\dagger}\rangle
+H_2\langle\chi\chi^{\dagger}\rangle
\label{lag42}
\;,
\eqa
where $L_i$ are the so-called low-energy constants ($i=0,1,2...10$), $H_i$
are the coefficients of the contact terms in chiral Lagrangian, and referred to as
high-energy constants ($i=1,2$).
The relations between the bare couplings $L_i$ and $H_i$ and their renormalized counterparts $L_i^r$ and $H_i^r$ are
\bqa
\label{Li}
L_i&=&L_i^r-{\Gamma_i\Lambda^{-2\epsilon}\over2(4\pi)^2}
\left[{1\over\epsilon}+1\right]\;,
\\
H_i&=&H_i^r-{\Delta_i\Lambda^{-2\epsilon}\over2(4\pi)^2}
\left[{1\over\epsilon}+1\right]\;.
\label{hi}
\eqa
The constants
$\Gamma_i$ { and $\Delta_i$}
assume the following values~\cite{gasser2}
\begin{align}
&  \Gamma_{4}=\frac{1}{8}\;,& \Gamma_{5}&=\frac{3}{8}\;,& \Gamma_{6}&={11\over144}\;,
  \\
& \Gamma_{7}=0\;,& \Gamma_{8}&={5\over48}\;,
  &\Delta_2&={5\over24} \;.
\end{align}
Since the bare parameters are independent of the scale $\Lambda$, differentiation
of Eqs.~(\ref{Li})--(\ref{hi}) 
immediately gives rise to equations governing the running of the
renormalized couplings. The renormalization group equations read
\bqa
\label{rl}
\Lambda{dL_i^r\over d\Lambda}=-{\Gamma_i\over(4\pi)^2}\;,
\hspace{0.6cm}\Lambda{dH_i^r\over d\Lambda}=-{\Delta_i\over(4\pi)^2}\;.
\eqa
We note that $\Gamma_7=0$, which implies that $L_7^r$ does not run and we write $L_7=L_7^r$.

\begin{widetext}
The quadratic part of the Lagrangian Eq.~(\ref{lag42}) is given by
\bqa\nonumber
{\cal L}_4^{\rm quadratic}&=&{4L_4\over f^2}(m_{\pi,0}^2+2m_{K,0}^2)
\left[2\partial_{\mu}\pi^+\partial^{\mu}\pi^-+\partial_{\mu}\pi^0\partial^{\mu}\pi^0
+2\partial_{\mu}K^+\partial^{\mu}K^-+2\partial_{\mu}K^0\partial^{\mu}\bar{K}^0
+\partial_{\mu}\eta\partial^{\mu}\eta\right]
\\ \nonumber
&&+{4L_5\over f^2}\left[m_{\pi,0}^2(2\partial_{\mu}\pi^+\partial^{\mu}\pi^-
+\partial_{\mu}\pi^0\partial^{\mu}\pi^0)
+2m_{K,0}^2(\partial_{\mu}K^+\partial^{\mu}K^-+
\partial_{\mu}K^0\partial^{\mu}\bar{K}^0)+
m_{\eta,0}^2\partial_{\mu}\eta\partial^{\mu}\eta
\right]
\\ \nonumber
&&-{8L_6\over f^2}(m_{\pi,0}^2+2m_{K,0}^2)
\left[m_{\pi,0}^2(2\pi^+\pi^-+(\pi^0)^2)+2m_{K,0}^2(K^+K^-+K^0\bar{K}^0)+
m_{\eta,0}^2\eta^2
\right]\\ \nonumber
&& 
-{64L_7\over3{f^2}}
(m_{\pi,0}^2-m_{K,0}^2)^2\eta^2
-{16L_8\over f^2}\left[
m_{\pi,0}^4\pi^+\pi^-+{1\over2}m_{\pi,0}^4(\pi^0)^2
+m_{K,0}^4(K^+K^-+K^0\bar{K}^0)
\right. \\ &&\left. 
+{1\over3}\left(4m_{K,0}^4
-4m_{\pi,0}^2m_{K,0}^2+{3\over2}m_{\pi,0}^4\right)\eta^2
\right]
\;.
\label{nloq}
\eqa
\end{widetext}
Finally, there are static terms from ${\cal L}_6$ that contribute at ${\cal O}(p^6)$
to the pressure, but they are temperature independent and only serve to renormalize
the vacuum energy.

\section{Pressure}
The free energy density is given by 
\bqa
{\cal F}&=&-{T\over V_{\rm sys}}\log{\cal Z}\;,
\eqa
where $V_{\rm sys}$ is the volume of the system and ${\cal Z}$ is the partition function which can be
expressed as a path integral in the imaginary-time formalism
\bqa
{\cal Z}&=&\int{\cal D}\phi e^{-\int_0^{\beta}d\tau\int d^3x{\cal L}_{E}}\;,
\eqa
where ${\cal L}_E$ is the Euclidean Lagrangian, $\beta\equiv 1/T$,
and $\phi$ is short-hand notation for all the fields integrated over. 
The pressure is then given by ${\cal P}=-{\cal F}$.
The loop diagrams that contribute to the pressure are ultraviolet divergent and must
be regularized. We use dimensional regularization where power divergences are set to zero
and logarithmic divergences show up as poles in $\epsilon$, where $d=3-2\epsilon$.
There are both temperature-independent and temperature-dependent divergences.
The counterterms diagrams that are used to cancel the $T=0$ divergences are also sufficient
to cancel the temperature-dependent ones. In the present paper, we are interested in finite-temperature effects and so we simply throw away the $T=0$ divergences.

The ${\cal O}(p^2)$ contribution is given by the static part of the Lagrangian ${\cal L}_2$.
Since this term is temperature independent, we ignore it henceforth.
In the following, we denote the finite-temperature
contribution at ${\cal O}(p^{2n})$ by ${\cal P}_{n-1}$
with $n=1,2,3..$. The result through ${\cal O}(p^{2n})$ is denoted
by ${\cal P}_{0+1+...n-1}$. 

\subsection{${\cal O}(p^4)$}
The one-loop pressure is given by 
\bqa
{\cal P}_1&=&{3\over2}I_0^{\prime}(m^2_{\pi,0})+2I_0^{\prime}(m^2_{K,0})
+{1\over2}I_0^{\prime}(m^2_{\eta,0})\;,
\eqa
where $I_0^{\prime}(m)$ is given by Eq.~(\ref{i0p}). Since we are only interested in the
temperature dependence, we keep the terms $J_0(\beta m)$ to obtain
\bqa
\nonumber
{\cal P}_{0+1}&=&{T^4\over(4\pi)^2}\left[{3\over2}J_0(\beta m_{\pi,0})+2J_0(\beta m_{K,0})
\right.\\ && \left.
+{1\over2}J_0(\beta m_{\eta,0})\right]\;,
\label{p1}
\eqa
{where the thermal integrals $J_n(\beta m)$ are defined in Eq.~(\ref{defj})
and} where $J_0(\beta m)$ is to be evaluated at $\epsilon=0$.

\subsection{${\cal O}(p^6)$}
At ${\cal O}(p^6)$, there are three contributions to the pressure: the tree-level
graphs, the one-loop graphs with a mass or derivative insertion, and the two-loop graphs.
The tree graphs are temperature independent and discarded. The one-loop diagrams
can be split into a temperature-independent term and a temperature-dependent term, where both
of them are divergent.
The two-loop graphs can be split in a similar manner. The temperature-dependent divergent parts
from the one-loop graphs cancel against the temperature-dependent divergent parts from
the two-loop graphs, showing that renormalization at $T=0$ is sufficient to obtain a 
finite expression for the pressure.

\begin{widetext}
The two-loop graphs are shown in the Fig.~\ref{graph0}. Their expression is
\bqa
{\cal P}_2^a&=&-{m_{\pi,0}^2\over f^2}
\left[{3\over8}I_1^2(m_{\pi,0}^2)
-{1\over4}I_1(m_{\pi,0}^2)I_1(m_{\eta,0}^2)
+{7\over72}I_1^2(m_{\eta,0}^2)
\right]
-{m_{K,0}^2\over f^2}\left[
{2\over3}I_1(m_{K,0}^2)I_1(m_{\eta,0}^2)
-{2\over9}I_1^2(m_{\eta,0}^2)
\right]\;,
\label{p2a}
\eqa
where the integral $I_1(m^2)$ is defined in Eq.~(\ref{i1}).
The one-loop counterterm graphs are shown in Fig.~\ref{graph1}. 
Their expression is
\bqa\nonumber
{\cal P}_2^b&=&{4L_4-8L_6\over f^2}(m^2_{\pi,0}+2m^2_{K,0})\left[
3m^2_{\pi,0}I_1(m^2_{\pi,0})
+4m^2_{K,0}I_1(m^2_{K,0})
+m^2_{\eta,0}I_1(m^2_{\eta,0})
\right]
\\ \nonumber&& 
+{4L_5\over f^2}\left[
3m^4_{\pi,0}I_1(m^2_{\pi,0})
+4m^4_{K,0}I_1(m^2_{K,0})
+m^4_{\eta,0}I_1(m^2_{\eta,0})
\right]
-{64L_7\over3{f^2}}(m_{\pi,0}^2-m_{K,0}^2)^2I_1(m_{\eta,0}^2)\\ &&
-{8L_8\over f^2}\left[
3m^4_{\pi,0}I_1(m^2_{\pi,0})
+4m^4_{K,0}I_1(m^2_{K,0})
+{{8m^4_{K,0}-8m^2_{K,0}m^2_{\pi,0}+3m^4_{\pi,0}}\over3}I_1(m^2_{\eta,0})\right]\;.
\label{p2b}
\eqa

\begin{figure}
\includegraphics[width=0.35\textwidth]{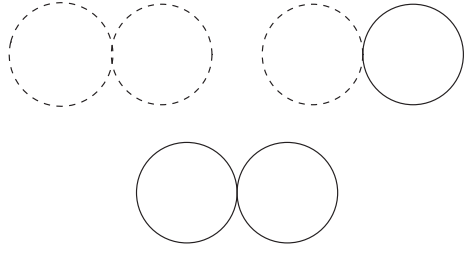}
\caption{
Two-loop Feynman graphs contributing to the pressure
at ${\cal O}(p^6)$.
Dashed line represents a neutral meson and solid line represents a charged meson.
}
\label{graph0}
\end{figure}

\begin{figure}
\includegraphics[width=0.2\textwidth]{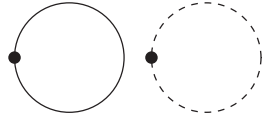}
\caption{
One-loop Feynman graphs with a mass or derivate counterterm insertion (indicated by a black blob)
contributing to the pressure at  ${\cal O}(p^6)$.
}
\label{graph1}
\end{figure}

Adding Eqs.~(\ref{p2a}) and (\ref{p2b}), and renormalizing the couplings using Eq.~(\ref{Li}),
we obtain 
\bqa\nonumber
{\cal P}_{2}&=&
-{m_{\pi,0}^2T^4\over(4\pi)^4f^2}\left[
{3\over8}J_1^2(\beta m_{\pi,0})
-{1\over4}J_1(\beta m_{\pi,0})J_1(\beta m_{\eta,0})
+{7\over72}J_1^2(\beta m_{\eta,0})
\right]
-{m_{K,0}^2T^4\over(4\pi)^4f^2}\left[
{2\over3}J_1(\beta m_{K,0})J_1(\beta m_{\eta,0})
\right.  \\&& \nonumber \left.
-{2\over9}J_1^2(\beta m_{\eta,0})
\right]
+{(4L_4^r-8L_6^r)T^2\over (4\pi)^2f^2}(m_{\pi,0}^2+2m_{K,0}^2)
\left[3m_{\pi,0}^2J_1(\beta m_{\pi,0})+
4m_{K,0}^2J_1(\beta m_{K,0})+m_{\eta,0}^2J_1(\beta m_{\eta,0})
\right]
\\ && \nonumber
+{4L_5^r T^2\over (4\pi)^2f^2}\left[
3m_{\pi,0}^4J_1(\beta m_{\pi,0})+4m_{K,0}^4J_1(\beta m_{K,0})+m_{\eta,0}^4J_1(\beta m_{\eta,0})
\right]
-{64L_7^r T^2\over3(4\pi)^2f^2}(m_{\pi,0}^2-m_{K,0}^2)^2J_1(\beta m_{\eta,0})
\\ && \nonumber
-{8L_8^r T^2\over (4\pi)^2f^2}\left[3m^4_{\pi,0}J_1(\beta m_{\pi,0})
+4m_{K,0}^4J_1(\beta m_{K,0})
+{1\over3}\left(8m_{K,0}^4
-8m_{\pi,0}^2m_{K,0}^2+3m_{\pi,0}^4\right)J_1(\beta m_{\eta,0})
\right]  \\ \nonumber&&  
+{T^2J_1(\beta m_{\pi,0})\over(4\pi)^4f^2}\left(
{3\over4}{m^4_{\pi,0}}\log{\Lambda^2\over m^2_{\pi,0}}
-{1\over4}{m^2_{\pi,0}m^2_{\eta,0}}
\log{\Lambda^2\over m^2_{\eta,0}}\right)
+{T^2J_1(\beta m_{K,0})\over(4\pi)^4f^2}\left(
{2\over3}{m^2_{K,0}m^2_{\eta,0}}\log{\Lambda^2\over m^2_{\eta,0}}\right)
\\ && 
+{T^2J_1(\beta m_{\eta,0})\over(4\pi)^4f^2}\left(
-{1\over4}{m^4_{\pi,0}}\log{\Lambda^2\over m^2_{\pi,0}}
+{2\over3}{m^4_{K,0}}\log{\Lambda^2\over m^2_{K,0}}
-{1\over3}{m^4_{\eta,0}}\log{\Lambda^2\over m^2_{\eta,0}}
+{1\over12}{m^2_{\pi,0}m^2_{\eta,0}}\log{\Lambda^2\over m^2_{\eta,0}}\right)\;,
\label{renp2}
\eqa
{where $J_1(\beta m)$ is to be evaluated at $\epsilon=0$.}

The terms proportional to the renormalized couplings $L_i^r$ 
{and the logarithms}
can be absorbed in
the one-loop result by replacing the bare meson masses with the physical meson masses at one loop,
listed in Appendix~\ref{mesonmass}. This can be seen by
writing the meson masses schematically as $m^2+\delta m^2$ and expanding
the one-loop contribution as
\bqa
I_0^{\prime}(m^2+\delta m^2)&=&I_0^{\prime}(m^2)-\delta m^2I_1(m^2)\;,
\eqa
where we have used Eq.~(\ref{reli}). Similarly, using Eq.~(\ref{recj}) for $\epsilon=0$,
we find
\bqa
J_0(\beta\sqrt{m^2+\delta m^2})
&=&J_0(\beta m)-\delta m^2\beta^2J_1(\beta m)\;.
\eqa
The sum of Eq.~(\ref{p1}) and~(\ref{renp2}) gives in the limit $\epsilon\rightarrow0$,
the finite-temperature pressure through ${\cal O}(p^6)$
\bqa\nonumber
{\cal P}_{0+1+2}&=&
{T^4\over(4\pi)^2}\left[{3\over2}J_0(\beta {M_{\pi}})+2J_0(\beta {M_{K}})
+{1\over2}J_0(\beta {M_{\eta}})\right]
-{m_{\pi,0}^2T^4\over(4\pi)^4f^2}\left[
{3\over8}J_1^2(\beta m_{\pi,0})
-{1\over4}J_1(\beta m_{\pi,0})J_1(\beta m_{\eta,0})
\right.\\ && \left.
+{7\over72}J_1^2(\beta m_{\eta,0})
\right]
-{m_{K,0}^2T^4\over(4\pi)^4f^2}\left[
{2\over3}J_1(\beta m_{K,0})J_1(\beta m_{\eta,0})
-{2\over9}J_1^2(\beta m_{\eta,0})
\right]\;.
\eqa
\end{widetext}
We note that the result simplifies significantly in the chiral limit since the terms
proportional to $m_{\pi,0}^2$ vanish. In the two-flavor case, the correction to the Stefan-Boltzmann result is of ${\cal O}(p^8)$, cf. Eq.~(\ref{p8}).

\section{Including electromagnetic interactions}
Electromagnetic interactions in the framework of chiral perturbation therory
were first included by Urech in Ref.~\cite{urech1} in the three-flavor case.
The $SU(2)_V$ symmetry of the chiral Lagrangian is then becoming a local $U(1)$
gauge symmetry.
Moreover, he showed that one can find a consistent power counting scheme also in this case, if 
the electromagnetic coupling $e$ counts as ${\cal O}(p)$ and the electromagnetic field $A_{\mu}$
counts as ${\cal O}(1)$.
The leading-order Lagrangian is now given by~\cite{e2chi}
\bqa\nonumber
{\cal L}_2&=&-{1\over4}F_{\mu\nu}F^{\mu\nu}+
{1\over4}f^2\langle\nabla_{\mu}\Sigma\nabla^{\mu}\Sigma^{\dagger}\rangle
\\ && \nonumber
+{1\over4}f^2\langle\chi^{\dagger}\Sigma+\Sigma^{\dagger}
\chi\rangle
+C\langle Q\Sigma Q\Sigma^{\dagger}\rangle
\\ &&
+{\cal L}_{\rm gf}+{\cal L}_{\rm ghost}
\label{newlag}
\;,
\eqa
where the first term is the kinetic term for the photons.
The covariant derivatives are
\bqa
\nabla_{\mu}\Sigma&=&\partial_{\mu}\Sigma+i[A_{\mu}Q,\Sigma]\;,
\\
\nabla_{\mu}\Sigma^{\dagger}&=&\partial_{\mu}\Sigma^{\dagger}+i[A_{\mu}Q,\Sigma^{\dagger}]\;.
\eqa
where the charge matrix of the quarks is 
\bqa
Q&=&
{1\over2}e\left(\lambda_3+{1\over\sqrt{3}}\lambda_8\right)
\;.
\eqa
Since our calculations involve the dynamical gauge field $A_{\mu}$, we need to fix the
gauge. In the class of covariant gauges,
the gauge-fixing term is
\bqa
{\cal L}_{\rm gf}&=&{1\over2\xi}(\partial_{\mu}A^{\mu})^2\;,
\eqa
where $\xi$ is the gauge-fixing parameter. The corresponding ghost term is
\bqa
{\cal L}_{\rm ghost}&=&
{1\over2}\partial_{\mu}\bar{c}\partial^{\mu}c\;.
\eqa
The ghost completely decouples from the rest of the Lagrangian.
In a general covariant gauge, the Euclidean space photon and ghost propagators are
\bqa
\Delta_{\mu\nu}(P)&=&{1\over P^2}\left(\delta_{\mu\nu}-(1-\xi){P_{\mu}P_{\nu}\over P^2}\right)\;,
\\
\Delta_{\rm ghost}(P)&=&{1\over P^2}\;.
\label{ghostprop}
\eqa
At ${\cal O}(p^4)$, the partial derivatives are also replaced by covariant derivatives in
Eq.~(\ref{lag42}).
The ${\cal O}(p^4)$ chiral Lagrangian has an additional 17 terms
whose coefficients were computed in the Feynman gauge, $\xi=1$~\cite{urech1,meis1,meis2,urech2}. 
Generally, the coefficients of the operators depend on the gauge, an explicit example is given 
in Ref.~\cite{gas}. Some of the operators have two powers of $e$ and two derivatives,
or two powers of $e$ with one power of the quark mass, or four powers of $e$.
The 14 operators required are
\bqa\nonumber
{\cal L}_4^Q&=&
K_1f^2\langle\nabla_{\mu}\Sigma^{\dagger}\nabla^{\mu}\Sigma\rangle\langle Q^2\rangle
\\ && \nonumber
+K_2f^2\langle\nabla_{\mu}\Sigma^{\dagger}\nabla^{\mu}\Sigma\rangle
\langle Q\Sigma Q\Sigma^{\dagger}\rangle+
\\ &&
\nonumber
+K_3f^2\left(\langle\nabla_{\mu}\Sigma^{\dagger}Q\Sigma\rangle
\langle\nabla^{\mu}\Sigma^{\dagger}Q\Sigma\rangle
+\langle\nabla_{\mu}\Sigma Q\Sigma^{\dagger}\rangle
\langle\nabla^{\mu}\Sigma Q\Sigma^{\dagger}\rangle
\right)
\\ && \nonumber
+K_4f^2\langle\nabla_{\mu}\Sigma^{\dagger}Q\Sigma\rangle
\langle\nabla^{\mu}\Sigma Q\Sigma^{\dagger}\rangle
\\ && \nonumber
+K_5f^2\langle(\nabla_{\mu}\Sigma^{\dagger}\nabla^{\mu}\Sigma+\nabla_{\mu}\Sigma\nabla^{\mu}\Sigma^{\dagger})Q^2\rangle
\\ && \nonumber
+K_6f^2\langle\nabla_{\mu}\Sigma^{\dagger}\nabla^{\mu}\Sigma
Q\Sigma^{\dagger} Q\Sigma
+\nabla_{\mu}\Sigma\nabla^{\mu}\Sigma^{\dagger}
Q\Sigma Q\Sigma^{\dagger}
\rangle
\\ &&
\nonumber
+K_7f^2\langle\chi^{\dagger}\Sigma+\Sigma^{\dagger}\chi\rangle
\langle Q^2\rangle
\\ \nonumber
&&
+K_8f^2\langle\chi^{\dagger}\Sigma+\Sigma^{\dagger}\chi\rangle
\langle Q\Sigma Q\Sigma^{\dagger}\rangle
\\ && \nonumber
+K_9f^2\langle(\chi^{\dagger}\Sigma+\Sigma^{\dagger}\chi
+\chi\Sigma^{\dagger}+\Sigma\chi^{\dagger})Q^2\rangle
\\ \nonumber
&&+K_{10}f^2\langle
(\chi^{\dagger}\Sigma+\Sigma^{\dagger}\chi)Q\Sigma^{\dagger}Q\Sigma
+(\chi\Sigma^{\dagger}+\Sigma\chi^{\dagger})Q\Sigma Q\Sigma^{\dagger}
\rangle
\\ && \nonumber
+K_{11}f^2\langle
(\chi^{\dagger}\Sigma-\Sigma^{\dagger}\chi)Q\Sigma^{\dagger}Q\Sigma
+(\chi\Sigma^{\dagger}-\Sigma\chi^{\dagger})Q\Sigma Q\Sigma^{\dagger}
\rangle
\\&& \nonumber
+K_{15}f^4\langle Q\Sigma Q\Sigma^{\dagger}\rangle^2
+K_{16}f^4\langle Q\Sigma Q\Sigma^{\dagger}\rangle\langle Q^2\rangle
\\ &&
+K_{17}f^4\langle Q^2\rangle^2\;,
\label{l4q}
\eqa
where $K_1$--$K_{17}$ are constants. The last operator is a contact term.
The relation between the bare and 
renormalized couplings is 
\bqa
\label{ki}
K_i&=&K_i^r-{\Lambda^{-2\epsilon}\Sigma_i\over2(4\pi)^2}\left[{1\over\epsilon}+1\right]\;,
\eqa
where the constants $\Sigma_i$ are
\begin{align}
   \label{LECs}
       \Sigma_1&={3\over4}\;,        \Sigma_2=Z\;,\\
        \Sigma_3&=-{3\over4}\;,         \Sigma_4=2Z\;,\\        
    \Sigma_5&=-{9\over4}\;,         \Sigma_6={3\over2}Z\;,\\
            \Sigma_7&=0\;,         \Sigma_8=Z\;,\\
    \Sigma_9&=-{1\over4}\;,         \Sigma_{10}={1\over4}+{3\over2}Z\;,\\
            \Sigma_{11}&={1\over8}\;,         \Sigma_{15}=
            {3\over2}+3Z+14Z^2
            \;,\\
    \Sigma_{16}&=-3-{3\over2}Z-Z^2\;,         \Sigma_{17}={3\over2}-{3\over2}Z+5Z^2\;,
\end{align}
and $Z={C\over f^4}$.
The running of $K_i^r$ is given by the solution to the renormalization group equation
\bqa
\Lambda{dK_i^r\over d\Lambda}&=&-{\Sigma_i\over(4\pi)^2}\;.
\label{rki}
\eqa
Note that $\Sigma_7=0$ which implies that 
$K_7$ does not run and we write $K_7=K_7^r$.

The charged mesons receive a contribution to the tree-level mass from 
{the term $C\langle Q\Sigma Q\Sigma^{\dagger}\rangle$ in 
the Lagrangian Eq.~(\ref{newlag}). Expanding this term to second order in the fields, we find
\bqa
\nonumber
\delta {\cal L}_2^{\rm quadratic}&=&-{Ce^2\over f^2}\left[
\phi_1^2+\phi_2^2+\phi_4^2+\phi_2^5
\right]\\
&=&
-2{Ce^2\over f^2}\left[
\pi^+\pi^-+K^+K^-\right]
\;,
\eqa
and therefore}
\bqa
\label{charge1}
m_{\pi^{\pm},0}^2&=&
B_0(m_u+m_d)
+2{Ce^2\over f^2}\;,\\
m_{K^{\pm},0}^2&=&
B_0(m_u+m_s)
+2{Ce^2\over f^2}\;.
\label{charge2}
\eqa
The new term which is of purely electromagnetic origin gives rise to the mass splitting of the
neutral and charged mesons that is nonzero in the chiral limit.

\begin{widetext} 
We also need the Lagrangian to fourth order in the fields. The new terms are coming from the covariant
derivative and from the term $C\langle Q\Sigma Q\Sigma^{\dagger}\rangle$. We find
\bqa\nonumber
{\cal L}_2^{Q,\rm quartic}&=&
{Ce^2\over6f^4}\left[
8(\pi^+\pi^-)^2+4\pi^+\pi^-(\pi^0)^2
+16\pi^+\pi^-K^+K^-
+2\pi^+\pi^-K^0\bar{K}^0
+(\pi^0)^2K^+K^-
+8(K^+K^-)^2
\right.\\ &&\nonumber \left.
+2K^+K^-K^0\bar{K}^0
+3K^+K^-\eta^2\right]
+ie(\pi^+\partial_{\mu}\pi^--\pi^-\partial_{\mu}\pi^+)A^{\mu}
+ie(K^+\partial_{\mu}K^--K^-\partial_{\mu}K^+)A^{\mu}
\\&& +e^2(\pi^+\pi^-+K^+K^-)A_{\mu}A^{\mu}
\;.
\label{quarticq}
\eqa
The one-loop counterterms are found by expanding 
${\cal L}_4^{Q}$ in Eq.~(\ref{l4q})
to second order in the fields. One finds
\bqa\nonumber
{\cal L}_4^{Q,\rm quadratic}&=&
{4\over3}e^2(K_1+K_2)\left[
\partial_{\mu}\pi^0\partial^{\mu}\pi^0+2\partial_{\mu}\pi^+\partial^{\mu}\pi^-
+2\partial_{\mu}K^+\partial^{\mu}K^-+2\partial_{\mu}K^0\partial^{\mu}\bar{K}^0+
\partial_{\mu}\eta\partial^{\mu}\eta
\right]
\\ \nonumber
&& 
-{1\over3}e^2(2K_3-K_4)[3\partial_{\mu}\pi^0\partial^{\mu}\pi^0
+\partial_{\mu}\eta\partial^{\mu}\eta]
\\ && \nonumber
+{2\over9}e^2(K_5+K_6)[5\partial_\mu\pi^0\partial^\mu \pi^0+10\partial_\mu\pi^+\partial^\mu\pi^-+10\partial_\mu K^+\partial^\mu K^-
+4\partial_\mu K^0\partial^\mu\bar{K}^0+3\partial_\mu\eta\partial^\mu\eta]
\\ \nonumber
&&-{4\over3}e^2(K_7+K_8)
\left[m_{\pi,0}^2(\pi^0)^2+2m_{\pi,0}^2\pi^+\pi^-
+2m_{K,0}^2(K^+K^-+K^0\bar{K}^0)+m_{\eta,0}^2\eta^2
\right]
\\ && \nonumber
-4e^2K_8(m_{\pi,0}^2+2m_{K,0}^2)(\pi^+\pi^-+K^+K^-)\\ && \nonumber
-\frac{2e^2 K_{9}}{27}\left[ m_{\pi,0}^2(30\pi^+\pi^- +15(\pi^0)^2 +18K^+K^- +\eta^2)
+m_{K,0}^2(12K^+K^- +12K^0\bar{K}^0 +8\eta^2) \right]\nonumber\\&&
-\frac{2e^{2}K_{10}}{27}\left[m_{\pi,0}^2(138\pi^+\pi^- +15(\pi^0)^2 +18K^+K^- +\eta^2)
+m_{K,0}^2(120K^+K^- +12K^0\bar{K}^0 +8\eta^2)\right]\nonumber\\&&
-8e^{2}K_{11}(m_{\pi,0}^2\pi^{+}\pi^{-}+m_{K,0}^2K^{+}K^{-})
-\frac{8}{3}f^{2}e^{4}K_{15}(\pi^{+}\pi^{-}+K^{+}K^{-})\nonumber\\&&
-\frac{4}{3}f^{2}e^{4} K_{16}(\pi^{+}\pi^{-}+K^{+}K^{-})
\;.
\label{quad2}
\eqa
\end{widetext}
Again there will be static terms from ${\cal L}_6^Q$ contributing to the 
renormalization of the vacuum energy and we will not need them.

\section{Pressure revisited}
In this section, we calculate the pressure through ${\cal O}(p^6)$ including electromagnetic
interactions. Since the neutral and charged mesons are no longer degenerate in masses,
we must express the pressure in terms of all the five different meson masses.
As mentioned before, the chiral Lagrangian including virtual photons is known only to
${\cal O}(p^4)$. It therefore not possible to renormalize the vacuum energy 
through ${\cal O}(p^6)$, but it is possible to renormalize the finite-temperature part since the
counterterms at the relevant order are given by the ${\cal O}(p^4)$ Lagrangian.

\subsection{${\cal O}(p^4)$}
Again the temperature-independent ${\cal O}(p^2)$-term is omitted.
The mesonic one-loop contribution to the pressure is the same as before, except that the
charged masses have changed according to Eqs.~(\ref{charge1})--(\ref{charge2}).
In addition, there is a contribution from the massless photons, giving
\bqa\nonumber
{\cal P}_1&=&{1\over2}I_0^{\prime}(m_{\pi,0}^2)+I_0^{\prime}(m_{\pi^{\pm},0}^2)
+I_0^{\prime}(m_{K^{\pm},0}^2)+I_0^{\prime}(m_{K,0}^2)
\\&&
+{1\over2}I_0^{\prime}(m_{\eta,0}^2)
+{1\over2}(d-1)I_0^{\prime}(0)
\;,
\label{p1foton1}
\eqa
where $d=3-2\epsilon$.
Omitting the temperature-independent divergent terms yields
in the limit $\epsilon\rightarrow0$
\bqa\nonumber
{\cal P}_1&=&{T^4\over(4\pi)^2}\left[{1\over2}J_0(\beta m_{\pi,0})+J_0(\beta m_{\pi^{\pm},0})
+J_0(\beta m_{K,0})
\right. \\  && \left. 
+J_0(\beta m_{K^{\pm},0})
+{1\over2}J_0(\beta m_{\eta,0})+J_0(0)\right]\;,
\label{p1foton}
\eqa
where $J_0(0)={16\pi^4\over45}$.

\subsection{${\cal O}(p^6)$}

\begin{figure}[htb]
\includegraphics[width=0.2\textwidth]{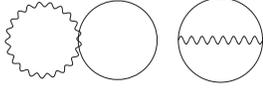}
\caption{
Feynman graphs contributing to the pressure
at next-to-next-to leading order.
Solid lines represent a charged meson and wavy lines represent a photon.
{The setting-sun diagram is shown to the right.}}
\label{graph3}
\end{figure}

The two-loop diagrams are those given in the previous section as well as a number of new
ones coming from the the interaction terms in Eq.~(\ref{quarticq}). The second group of diagrams
are shown in Fig.~\ref{graph3}. These are the only diagrams involving the photon 
propagator. We note in passing that the individual diagrams are gauge-fixing dependent, but the
sum is independent of $\xi$ in covariant gauge. The same result is obtained in 
the Coulomb gauge with gauge parameter $\xi$.

The expression for the diagrams in Fig.~\ref{graph0} is
\begin{widetext}
\bqa\nonumber
{\cal P}_2^a&=&-{m_{\pi,0}^2\over f^2}
\left[-{1\over8}I_1^2(m_{\pi,0}^2)
+{1\over6}I_1(m_{\pi^{\pm},0}^2)I_1(m_{\pi,0}^2)
-{1\over3}I_1^2(m_{\pi^{\pm},0}^2)
-{1\over6}I_1(m_{\pi^{\pm},0}^2)I_1(m_{K,0}^2)
\right. \\ \nonumber&&\left.
-{1\over6}I_1(m_{\pi^{\pm},0}^2)I_1(m_{K^{\pm},0}^2)
-{1\over12}I_1(m_{\pi,0}^2)I_1(m_{\eta,0}^2)
-{1\over6}I_1(m_{\pi^{\pm},0}^2)I_1(m_{\eta,0}^2)
+{7\over72}I_1^2(m_{\eta,0}^2)
\right]\\ \nonumber
&&-{m_{\pi^\pm,0}^2\over f^2}
\left[{1\over3}I_1(m_{\pi^{\pm},0}^2)I_1(m_{\pi,0}^2)
+{1\over3}I_1^2(m_{\pi^{\pm},0}^2)
+{1\over6}I_1(m_{\pi^{\pm},0}^2)I_1(m_{K,0}^2)
+{1\over6}I_1(m_{\pi^{\pm},0}^2)I_1(m_{K^{\pm},0}^2)\right]
\\ \nonumber
&&-{m_{K,0}^2\over f^2}\left[-{1\over3}I_1^2(m_{K^{\pm},0}^2)
-{1\over12}I_1(m_{\pi,0}^2)I_1(m_{K^{\pm},0}^2)
-{1\over6}I_1(m_{\pi^{\pm},0}^2)I_1(m_{K^{\pm},0}^2)
-{1\over6}I_1(m_{K,0}^2)I_1(m_{K^{\pm},0}^2)
\right. \\ \nonumber&&\left.
+{1\over3}I_1(m_{K^0,0}^2)I_1(m_{\eta,0}^2)
+{1\over12}I_1(m_{K^\pm,0}^2)I_1(m_{\eta,0}^2)
-{2\over9}I_1^2(m_{\eta,0}^2)
\right]\\ \nonumber
&&
-{m_{K^\pm,0}^2\over f^2}
\left[{1\over3}I_1^2(m_{K^{\pm},0}^2)+{1\over12}I_1(m_{\pi,0}^2)I_1(m_{K^{\pm},0}^2)
+{1\over6}I_1(m_{\pi^{\pm},0}^2)I_1(m_{K^{\pm},0}^2)
\right. \\ &&\left.
+{1\over6}I_1(m_{K,0}^2)I_1(m_{K^{\pm},0}^2)
+{1\over4}I_1(m_{K^\pm,0}^2)I_1(m_{\eta,0}^2)\right]
\;,
\label{p2aprime}
\eqa
where the charged masses are given by Eqs.~(\ref{charge1})--(\ref{charge2}).
Setting $e=0$, i.e. for degenerate meson masses, Eq.~(\ref{p2aprime}) reduces
to Eq.~(\ref{p2a}), as it should.

The first set of one-loop graphs with insertions is 
{shown in Fig.~\ref{graph1}.}
Their expression is
\bqa\nonumber
{\cal P}_2^b&=&
{4L_4\over f^2}(m_{\pi,0}^2+2m_{K,0}^2)
\left[2m_{\pi^{\pm},0}^2I_1(m^2_{\pi^{\pm},0})+m_{\pi,0}^2I_1(m^2_{\pi,0})+
2m_{K^{\pm},0}^2I_1(m^2_{K^{\pm},0})
+2m_{K,0}^2I_1(m^2_{K,0})+m_{\eta,0}^2I_1(m^2_{\eta,0})
\right]
\\ && \nonumber
+{4L_5\over f^2}\left[
m_{\pi,0}^2\left(2m_{\pi^{\pm},0}^2I_1(m^2_{\pi^{\pm},0})+m_{\pi,0}^2I_1(m^2_{\pi,0})\right)
+2m_{K,0}^2\left(m_{K^{\pm},0}^2I_1(m^2_{K^{\pm},0})
+m_{K,0}^2I_1(m^2_{K,0})\right)+m_{\eta,0}^4I_1(m^2_{\eta,0})
\right]
\\ && \nonumber
-{8L_6\over f^2}(m_{\pi,0}^2+2m_{K,0}^2)
\left[m_{\pi,0}^2\left(2I_1(m^2_{\pi^{\pm},0})+I_1(m^2_{\pi,0})\right)+
2m_{K,0}^2\left(I_1(m^2_{K^{\pm},0})
+I_1(m^2_{K,0})\right)+m_{\eta,0}^2I_1(m^2_{\eta,0})
\right]
\\ && \nonumber
-{64L_7 \over3f^2}(m_{\pi,0}^2-m_{K,0}^2)^2I_1(m^2_{\eta,0})
-{16L_8\over f^2}\left[m^4_{\pi,0}\left(I_1(m^2_{\pi^{\pm},0})+
{1\over2}I_1(m^2_{\pi,0})\right)
+m_{K,0}^4\left(I_1(m^2_{K^{\pm},0})+I_1(m^2_{K,0})\right)
\right. \\ &&\left.
+{1\over3}(4m_{K,0}^4
-4m_{\pi,0}^2m_{K,0}^2+{3\over2}m_{\pi,0}^4)I_1(m^2_{\eta,0})
\right]\;.
\label{bidrag2}
\eqa
The expression for diagrams arising from the interactions in Eq.~(\ref{quarticq}) {and
shown in Fig.~\ref{graph3}} is
\bqa
\nonumber
{\cal P}_2^c&=&
-(d-1)e^2I_1(m_{\pi^{\pm},0}^2)I_1(0)-{1\over2}e^2I_1^2(m_{\pi^{\pm},0}^2)
-2e^2m_{\pi^{\pm},0}^2I_{\rm sun}(m_{\pi^{\pm},0}^2)
-(d-1)
e^2I_1(m_{K^{\pm},0}^2)I_1(0)
\\ && \nonumber
-{1\over2}e^2I_1^2(m_{K^{\pm},0}^2)
-2e^2m_{K^{\pm},0}^2I_{\rm sun}(m_{K^{\pm},0}^2)
+{Ce^2\over6f^4}\left[
4I_1(m_{\pi,0}^2)I_1(m^2_{\pi^{\pm},0})
+16I_1^2(m_{\pi^{\pm},0}^2)
\right.\\ &&\nonumber \left.
+16I_1(m_{\pi^{\pm},0}^2)I_1(m_{K^{\pm},0}^2)
+I_1(m_{\pi,0}^2)I_1(m_{K^{\pm},0}^2)
+2I_1(m_{\pi^{\pm},0}^2)I_1(m_{K,0}^2)
+16I_1^2(m_{K^{\pm},0}^2)
\right.\\ &&\nonumber \left.
+2I_1(m_{K^{\pm},0}^2)I_1(m_{K,0}^2)
+3I_1(m_{K^{\pm},0}^2)I_1(m_{\eta,0}^2)
\right]\;,
\\ &&
\label{bidrag3}
\eqa
where $d=3-2\epsilon$ and $I_{\rm sun}(m^2)$ is defined in Eq.~(\ref{sundeffie}). 
{$I_{\rm sun}(m^2)$ is evaluated in Appendix \ref{suncalc}}.
Finally, the expression for the diagrams arising from Eq.~(\ref{quad2}) are given by
\bqa\nonumber
{\cal P}_2^d&=&
{4\over3}e^2(K_1+K_2)\left[
m^2_{\pi,0}I_1(m_{\pi,0}^2)+2m^2_{\pi^{\pm},0}I_1(m^2_{\pi^{\pm},0})+
2m^2_{K,0}I_1(m^2_{K,0})+2m^2_{K^{\pm},0}I_1(m^2_{K^{\pm},0})+
m^2_{\eta,0}I_1(m^2_{\eta,0})
\right]\\ \nonumber
&&+{1\over3}e^2(-2K_3+K_4)\left[3m^2_{\pi,0}I^2_1(m_{\pi,0})+m^2_{\eta,0}I^2_1(m_{\eta,0})\right]
+{2\over9}e^2(K_5+K_6)\left[
5m_{\pi,0}^2I_1(m_{\pi,0}^2)+10m_{\pi^{\pm},0}^2I_1(m_{\pi^{\pm},0}^2)
\right.\\ &&\nonumber \left.
+10m_{K^{\pm},0}^2I_1(m_{K^{\pm},0}^2)+4m_{K,0}^2I_1(m_{K,0}^2)
+3m_{\eta,0}^2I_1(m_{\eta,0}^2)
\right]
-{4\over3}e^2(K_7+K_8)\left[
m_{\pi,0}^2\left(I_1(m_{\pi,0}^2)+2I_1(m_{\pi^{\pm},0}^2)\right)
\right.\\ &&\nonumber \left.
+2m^2_{K,0}\left(I_1(m_{K^{\pm},0}^2)+I_1(m_{K,0}^2)\right)
+m_{\eta,0}^2I_1(m_{\eta,0}^2)
\right]
-4e^2K_8(m_{\pi,0}^2+2m_{K,0}^2)\left[I_1(m_{\pi^{\pm},0}^2)+I_1(m_{K^{\pm},0}^2)\right]
\\ && \nonumber
-\frac{2e^2 K_{9}}{27}\left[ m_{\pi,0}^2\left(30I_1(m_{\pi^{\pm},0}^2) +15I_1(m_{\pi,0}^2) +18I_1(m_{K^{\pm},0}^2) 
+I_1(m_{\eta,0}^2)\right)
+m_{K,0}^2\left(12I_1(m_{K^{\pm},0}^2) +12I_1(m_{K,0}^2) 
\right.\right.\\ && \nonumber\left.\left.
+8I_1(m_{\eta,0}^2)\right) \right]
-\frac{2e^2K_{10}}{27}\left[m_{\pi,0}^2\left(138I_1(m_{\pi^{\pm},0}^2) +15I_1(m_{\pi,0}^2) 
+18I_1(m_{K^{\pm},0}^2) +I_1(m_{\eta,0}^2)\right)
\right.\\ && 
\left.
+m_{K,0}^2\left(120I_1(m^2_{K^\pm,0})
+12I_1(m_{K,0}^2) +8I_1(m_{\eta,0}^2)\right)\right]
-8e^2K_{11}\left[m_{\pi,0}^2I_1(m_{\pi^{\pm},0}^2)+m_{K,0}^2I_1(m_{K^{\pm},0}^2)\right]\;.
\label{bidrag4}
\eqa
The complete result for the pressure is then given by the sum of 
Eqs.~(\ref{p1foton}),~(\ref{p2aprime}),~(\ref{bidrag2}), (\ref{bidrag3}), and ({\ref{bidrag4}}).
Again we can absorb the terms that involve the low-energy constants by replacing the
bare meson masses with their one-loop expression.
The final result is
\bqa\nonumber
{\cal P}_{0+1+2}&=&
{T^4\over(4\pi)^2}\left[{1\over2}J_0(\beta m_{\pi^{0}})+J_0(\beta m_{\pi^{\pm}})
+J_0(\beta m_{K^{0}})+J_0(\beta m_{K^{\pm}})
+{1\over2}J_0(\beta m_{\eta})+J_0(0)\right]
\\ \nonumber &&
-{m_{\pi,0}^2T^4\over (4\pi)^4f^2}\left[
{1\over2}J_1(\beta m_{\pi^{\pm},0})J_1(\beta m_{\pi,0})
-{1\over8}J_1^2(\beta m_{\pi,0})
-{1\over12}J_1(\beta m_{\pi,0})J_1(\beta m_{\eta,0})-{1\over6}J_1(\beta m_{\pi^{\pm},0})J_1(\beta m_{\eta,0})
\right.\\ && \nonumber\left.
+{7\over72}J_1^2(\beta m_{\eta,0})\right]
-{m_{K,0}^2T^4\over (4\pi)^4f^2}\left[
{1\over3}J_1(\beta m_{K,0})J_1(\beta m_{\eta,0})
+{1\over3}J_1(\beta m_{K^{\pm},0})J_1(\beta m_{\eta,0})
-{2\over9}J_1^2(\beta m_{\eta,0})\right]\\ \nonumber&&
-{e^2T^4\over (4\pi)^4}\left[2J_1(\beta m_{\pi^{\pm},0})J_1(0)+{1\over2}J_1^2(\beta m_{\pi^{\pm},0})+
2J_1(\beta m_{K^{\pm},0})J_1(0)+{1\over2}J_1^2(\beta m_{K^{\pm},0})\right]
\\ &&\nonumber
-{2m_{\pi^{\pm},0}^2e^2I^{(2)}_{\rm sun}(m_{\pi^{\pm},0}^2)}
-2m_{K^{\pm},0}^2e^2I^{(2)}_{\rm sun}(m_{K^{\pm},0}^2)
+{Ce^2T^4\over(4\pi)^4f^4}\left[2J_1^2(\beta m_{\pi^{\pm},0})
\right.\\ && 
\left.
+2J_1(\beta m_{\pi^{\pm},0})J_1(\beta m_{K^{\pm},0})+2J_1^2(\beta m_{K^{\pm},0})\right]\;.
\label{final2}
\eqa
\end{widetext}
where $I_{\rm sun}^{(2)}(m^2)$ is defined in Eq.~(\ref{2thermal}) and
we note that $J_1(0)={4\pi^2\over3}$.

\section{Quark condensates}
In the vacuum, the light and $s$-quark condensates
are defined as 
\bqa
\langle \bar{u}u\rangle_0&=&{\partial V\over\partial m_u}\;,\\
\langle \bar{d}d\rangle_0&=&{\partial V\over\partial m_d}\;,\\
\langle \bar{s}s\rangle_0&=&{\partial V\over\partial m_s}\;,
\eqa
where $V$ is the vacuum energy density. 
By introducing the sum $m={1\over2}(m_u+m_d)$ and difference $\Delta m={1\over2}(m_u-m_d)$
of the light quark masses, we calculate the sum and difference of the light quark condensates
as
\bqa
\langle \bar{u}u\rangle_0+\langle \bar{d}d\rangle_0&=&\langle \bar{q}q\rangle_0
={\partial V\over\partial m}\;,\\
\langle \bar{u}u\rangle_0-\langle \bar{d}d\rangle_0
&=&{\partial V\over\partial\Delta m}\;.
\eqa
At finite temperature, we replace $V$ by $V-{\cal P}$~\cite{finitet} and 
we therefore have
\bqa
\langle\bar{q}q\rangle&=&\langle \bar{q}q\rangle_0\left[1+
\sum_a{c_{a}\over f^2}
{\partial{\cal P}\over\partial m^2_a}
\right]\;,\\
\langle \bar{s}s\rangle&=&
\langle\bar{s}s\rangle_0
\left[1+\sum_a{c_{sa}\over f^2}
{\partial{\cal P}\over\partial m^2_{a}}
\right]\;,
\eqa
where the sum is over the eight mesons and 
the coefficients are
\bqa
c_{a}&=&-f^{2}{\partial m_{a}^2\over\partial m}\langle \bar{q}q\rangle_0^{-1}\;,\\
c_{sa}&=&-f^{2}{\partial m_{a}^2\over\partial m_s}\langle \bar{s}s\rangle_0^{-1}\;.
\eqa
The expressions for the coefficients are obtained by using the results for
the condensates at $T=0$ given by (\ref{light1})--(\ref{slast}) and
the meson masses listed in Eqs~(\ref{pi00})--(\ref{mes5}).

\section{Hadron resonance gas model}
In the hadron resonance gas (HRG) model, one approximates the partition function of the system
by the partition function of a gas of non-interacting hadrons and resonances.
The pressure ${\cal P}$ is therefore given by the sum of independent contributions 
${\cal P}_h$
coming from the different species,
\bqa
\nonumber
{\cal P}&=&\sum_{h}{\cal P}_h
\nonumber\\ \nonumber
&=&\mp{8T\over(4\pi)^2}\sum_hd_h(2s+1)
\int_0^{\infty}dp\,p^2\log\left[1\mp e^{-\beta\sqrt{p^2+m_h^2}}\right]\;,
\\ &&
\label{sumhad}
\eqa
where $d_h$ is the multiplicity, $s$ is the spin, $m_h$ is the hadron mass,
and the upper (lower) sign is for mesons (baryons).
The lightest hadrons we include in the sum are shown in Table~\ref{hadtable}.
{In the numerical work, the used HRG model includes more than 200 known mesons and baryons below 2.5 GeV in Particle Data Group~\cite{PDG}. As known from Ref.~\cite{tc3}, it is reasonable to add those known resonances. Of course, it includes those broad light flavor mesons e.g. $f_0(500)$, $f_0(1370)$, and $K^*_0(700)$ where we take the central values of the estimated masses.}

\begin{table}
\begin{tabular}{|l|c|c|c||l|c|c|c|c|}
\hline
hadron & $m (\textmd{MeV})$  & $s$ & $d_h$ & hadron & $m (\textmd{MeV})$ 
 & $s$ & $d_h$ \\ 
\hline
\hline
$\pi^{\pm}$ & 139.57  & 0 & 2  & $p$ & 938.27  & 1/2 & 2  \\
$\pi^0$ & 134.98  & 0 & 1 & $n$ & 939.57  & 1/2 & 2  \\
$K^{\pm}$ & 493.68  & 0 & 2 & $\eta'$ & 957.78  & 0 & 1  \\
$K^0/\bar{K}^0$ & 497.61  & 0 & 2 & $f_0$ & 990$\pm$20  & 0 & 1  \\
$\eta$ & 547.86  & 0 & 1 & $a_0$ & 980$\pm$20  & 1 & 1  \\ 
$\rho^{\pm}$ & 775.26  & 1 & 2 & $\phi$ & 1019.46  & 1 & 1  \\
$\rho$ & 775.26  & 1 & 1 & $\Lambda$ & 1115.68  & 1/2 & 1 \\  
$\omega$ & 782.66  & 1 & 1 & $h_1$ & 1166$\pm$6  & 1 & 1 \\ 
$K_{*}^{\pm}$ & 891.67  & 1 & 2 & $\Sigma^\pm$ & 1189.37  & 1/2 & 2 \\ 
$K_*^0$ & 895.55  & 1 & 2 &  $\Sigma^0$ & 1192.64  & 1/2 & 1 \\ 
\hline
\end{tabular}
\caption{Lightest hadrons included in the Hadron Resonance Gas model.}
\label{hadtable}
\end{table}

We also need the expressions for the condensates in the HRG model.
They are given by 
\bqa\nonumber
\langle\bar{q}q\rangle&=&\langle\bar{q}q\rangle_0-{\partial{\cal P}\over\partial m}\\
&=&\langle\bar{q}q\rangle_0+\sum_{h} n_h(T){\partial m_h\over\partial m}\;,
\\
\langle\bar{s}s\rangle&=&\langle\bar{s}s\rangle_0-{\partial{\cal P}\over\partial m_s}\\
&=&\langle\bar{s}s\rangle_0+\sum_{h} n_h(T){\partial m_h\over\partial m_s}\;, 
\eqa
where the temperature dependent density of hadrons is 
\bqa
\nonumber
n_h(T)=
{8d_h(2s+1)\over(4\pi)^2}
\int_0^{\infty}dp\,{m_hp^2\over\sqrt{p^2+m^2_h}}{1\over e^{\beta\sqrt{p^2+m_h^2}}\mp 1 }\;. 
\\ &&
\eqa
The derivatives of the hadrons masses with respect to the light quark mass $m$ and the strange quark mass $m_s$ can be written as~\cite{tc3}
\bqa
{\partial m_h\over\partial m}
&=&{2B_0}{\sigma_{\pi,h}\over m^2_{\pi^0}}\;,\\
{\partial m_h\over\partial m_s}&=&{\sigma_{s,h}\over m_s}
={\sigma_{s,h}\over m^2_{K,0}}{B_0(m+m_s)\over m_s}\;.
\eqa
The sigma terms for the fundamental states 
are taken from \cite{MartinCamalich:2010fp}. It is difficult to calculate the sigma terms for each particle, but we follow Ref. \cite{tc3} and assume that all hadrons have the same sigma term as their fundamental state. 

\section{Numerical results and discussion}
In this section, we present and discuss our numerical results.
As input we will use the physical meson masses and the pion decay constant
taken from the Particle Data Group~\cite{PDG}
\bqa
m_{\pi^0}&=&134.98\,{\rm MeV}\;, 
\\
m_{\pi^{\pm}}&=&139.57\,{\rm MeV}\;,
\\
m_{K^{\pm}}&=&493.68\,{\rm MeV}\;,
\\
m_{K^0}&=&497.61\,{\rm MeV}
\;,\\
m_{\eta}&=&547.86\,{\rm MeV}
\;,\\
f_{\pi}&=&92.07\,{\rm MeV}\;.
\eqa
The numerical values of the low-energy constants that we need 
are~\cite{bijnensreview,jamin},
where Ref.~\cite{bijnensreview} includes $L^r_{4}-L^r_{8}$ and $H^r_{2}$
is taken from Ref.~\cite{jamin},
\bqa
        {L}_{4}^r&=(0.0 \pm 0.3)\times10^{-3}\;,\\
        {L}_{5}^r&=(1.2 \pm 0.1)\times10^{-3}\;,\\
        {L}_{6}^r&=(0.0 \pm 0.4)\times10^{-3}\;,\\
        {L}_{7}^r&=(-0.3 \pm 0.2)\times10^{-3}\;,\\
        {L}_{8}^r&=(0.5 \pm 0.2)\times10^{-3}\;,\\
        {H}_{2}^r&=(-3.4\pm 1.5)\times10^{-3}\;.
\eqa
These couplings are at the scale of the $\rho$ mass, $\Lambda=775.26$ MeV.

Including electromagnetic interactions, we need a number of additional couplings.
The electromagnetic coupling is~\cite{PDG}
\bqa
e^2&=& 0.092
\;.
\eqa
The numerical value of the constant $C$ has been estimated by Urech~\cite{urech1}. Its value is
\bqa
C&=&61.1\times10^{-6}({\rm GeV})^4\;.
\eqa 
At tree level, this gives rise to a mass splitting between the neutral and charged
pion of approximately $4.8$ MeV, which is very close the experimental value of $4.6$ MeV.
Finally, we need~\cite{bijnensreview,Bijnens:1996kk,Gorghetto:2018ocs}
\bqa
    \label{LEC2s}
    {K}_{1}^r&=-2.7\times10^{-3}\;,\\
    {K}_{2}^r&=0.7\times10^{-3}\;,\\
    {K}_{3}^r&=2.7\times10^{-3}\;,\\
    {K}_{4}^r&=1.4\times10^{-3}\;,\\
    {K}_{5}^r&=11.6\times10^{-3}\;,\\
    {K}_{6}^r&=2.8\times10^{-3}\;,\\
    {K}_{7}^r&=0\times10^{-3}\;,\\
    {K}_{8}^r&=0\times10^{-3}\;,\\
    {K}_{9}^r&=-1.3\times10^{-3}\;,\\
    {K}_{10}^r&=4\times10^{-3}\;,\\
    {K}_{11}^r&=1.3\times10^{-3}\;,
\eqa    
where all the $K^r_i$ are assigned a conservative $100\%$ uncertainty \cite{Gorghetto:2018ocs}.
The low-energy constants 
$K^r_{1}-K^r_{8}$ and $K^r_{11}$ can be found in Ref.~\cite{bijnensreview}, $K^r_{9}$ and $K_{10}^r$ are from Ref.~\cite{Bijnens:1996kk}.

If we ignore electromagnetic interactions, the charged pion is degenerate with the
neutral pion, and the charged kaon is degenerate with the neutral kaon.
In this case, we use the experimental values for the masses of the neutral mesons
as well as $f_{\pi}$. Using these values
together with the low-energy constants, Eqs.~(\ref{pi00}), ~(\ref{k00}), and~(\ref{fpi})  
(here with $e=0$)
give us the tree-level values for 
$m_{\pi,0}$, $m_{K,0}$, and $f$. The tree-level value of the eta mass is then given by
the relation $m_{\eta,0}^2={1\over3}(4m_{K,0}^2-m_{\pi,0}^2)$.
The bare values we find are
\bqa
m_{\pi,0}&=&135.52\,{\rm MeV}\;, 
\\
m_{K,0}&=&536.72\,{\rm MeV}\;,
\\
m_{\eta,0}&=&614.79\,{\rm MeV}
\;,
\\
f&=&76.93\,{\rm MeV}\;.
\eqa
Adding electromagnetic effects, we obtain
\bqa
m_{\pi,0}&=&{135.97}\,{\rm MeV}\;, 
\\
m_{\pi^{\pm},0}&=& {137.11}\,{\rm MeV}\;,
\\
m_{K^{\pm},0}&=& {531.85}\,{\rm MeV}\;,
\\
m_{K,0}&=& {537.14}\,{\rm MeV}\;,
\\
m_{\eta,0}&=& {615.25}\,{\rm MeV}
\;,
\\
f&=& {76.69}\,{\rm MeV}\;.
\eqa
In both cases, we see that renormalization effects are modest, except for the pion-decay constant.

\begin{figure}[htb]
\centering
\includegraphics[width=0.48\textwidth]{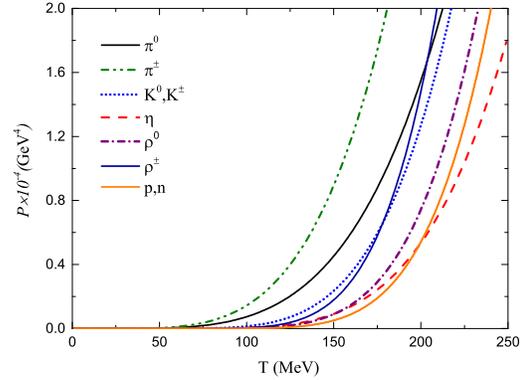}
\caption{Individual contributions to the pressure in the HRG model
as a function of the temperature in MeV. See main text for details.}
\label{pressureM}
\end{figure}
In Fig.~\ref{pressureM}, we show some of the 
the individual contributions to the pressure in the HRG model in units of $10^{-4}$ GeV$^4$
as a function of temperature in MeV. This is essentially the same as Fig.~2 of
Ref.~\cite{endrodihrg}. As expected, at any given temperature, the lightest states
contribute more to the total pressure than the heavier states. Up to approximately 100 MeV,
only the pions contribute significantly. From 120-130 MeV onwards, heavier states that are not
included in three-flavor $\chi$PT start to contribute significantly.

\begin{figure}[htb]
\centering
\includegraphics[width=0.48\textwidth]{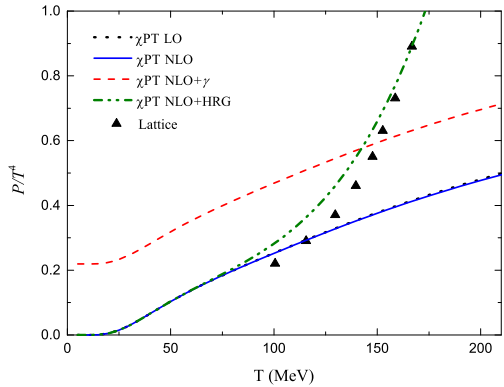}
\caption{Pressure normalized by $T^4$ as a function of the temperature
in MeV. See main text for details.}
\label{pressureM3}
\end{figure}
In Fig.~\ref{pressureM3}, we show the pressure ${\cal P}$ normalized to $T^4$ in various
approximations. The dotted line is the ${\cal O}(p^2)$ result in $\chi$PT, while
the blue line is  ${\cal O}(p^4)$ result. The red dashed line is the  ${\cal O}(p^4)$ result 
including electromagnetic effects. The green line shows the resulting normalized pressure
combining $\chi$PT and the Hadron Resonance Gas model.
{Doing this, the eight mesons in three-flavor $\chi$PT are excluded from the
sum in Eq.~(\ref{sumhad}) so we do not count degrees of freedom twice.}
The black triangles are the lattice results taken from 
Ref.~\cite{tc3}. For low temperatures, the contribution to the pressure from the
massive states is Boltzmann suppressed. This implies that the normalized pressure
vanishes, except for the case where the contribution from the photons is included.
The normalized pressure in the limit $T\rightarrow0$ is therefore 
$2\times{\pi^2\over90}=0.22$. The difference between the red and the blue line is fairly
constant over the temperature range shown, indicating that electromagnetic interactions
contribute relatively little to the total pressure. The green and blue lines are
essentially on top of each other until a temperature of approximately 90 MeV, 
where they start to deviate. The steep increase of the green curves shows the effects
of including heavier states. The agreement between the resulting normalized
pressure and the lattice result up to the largest temperatures is good.

\begin{figure}[htb]
\centering
\includegraphics[width=0.48\textwidth]{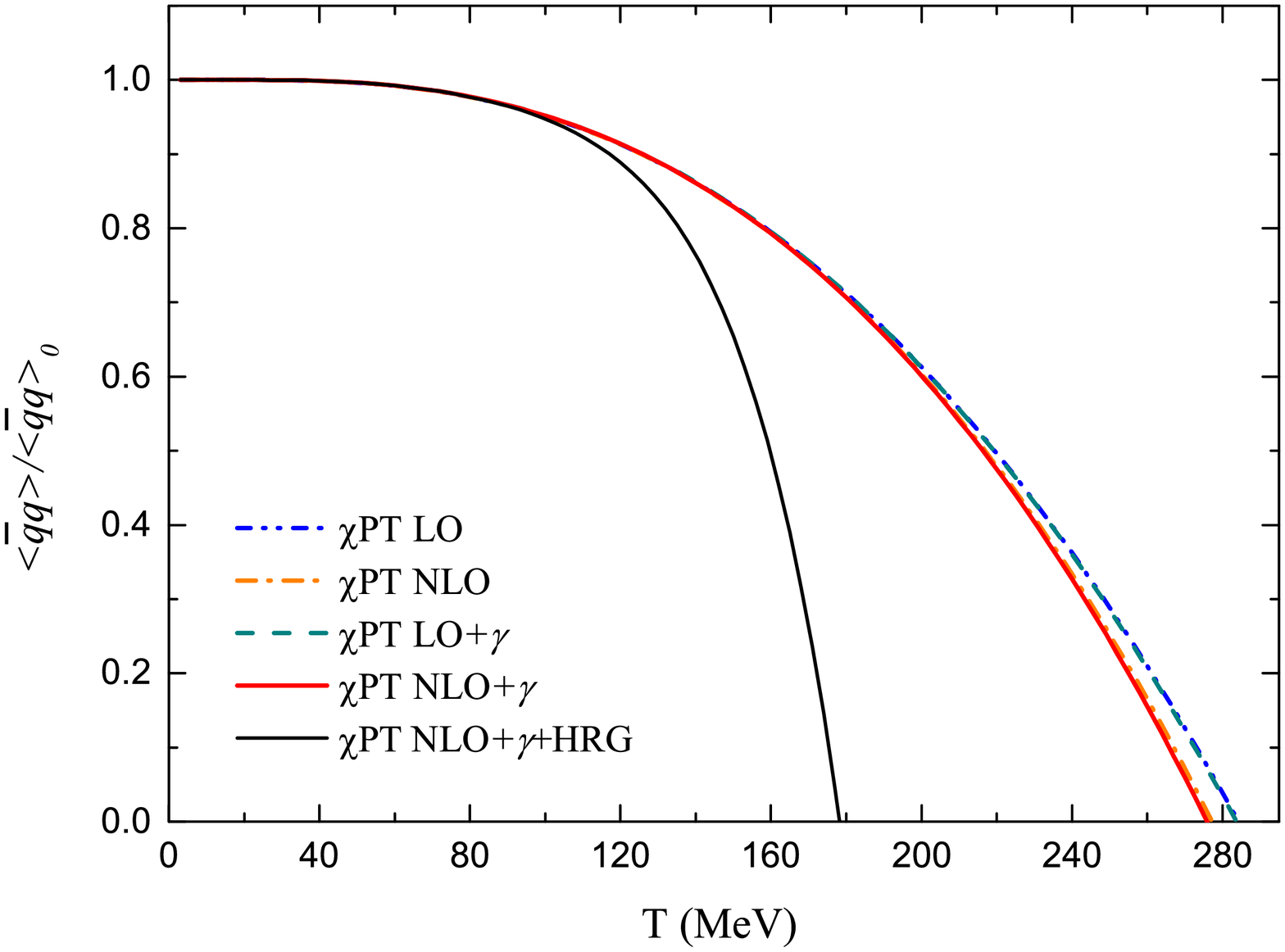}
\caption{Normalized light quark condensate  $\langle\bar{q}q\rangle/\langle\bar{q}q\rangle_0$ as a function of the temperature
in MeV. See main text for details.}
\label{qqbar}
\end{figure}

In Fig.~\ref{qqbar}, we show the light quark condensate normalized to its zero-temperature
value in different approximations as a function of the temperature. The blue (green) line is the ${\cal O}(p^2)$ result
without (with) electromagnetic interactions. The yellow (red) line is the ${\cal O}(p^4)$ result
without (with) electromagnetic interactions. For comparison, we show in black the result
from the HRG model. Obviously, $\chi$PT is not valid in the entire temperature range shown,
but for low temperatures, up to $T\simeq150$ MeV, it seems to be converging very well and 
electromagnetic effects are not very large.
{However, from Figs.~\ref{pressureM3} and~\ref{lhs} below, it is also clear
that $\chi$PT alone cannot explain the lattice results beyond approximately 120 MeV.}

\begin{figure}[htb]
\centering
\includegraphics[width=0.48\textwidth]{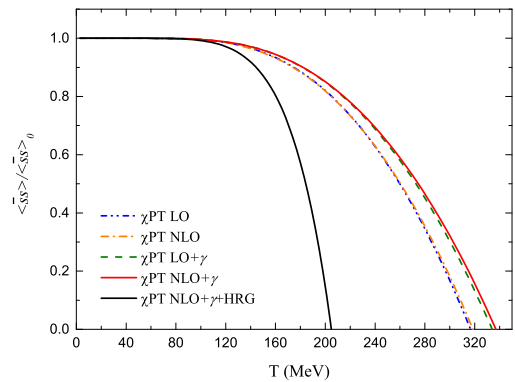}
\caption{Normalized strange quark condensate  $\langle\bar{s}s\rangle/\langle\bar{s}s\rangle_0$ as a function of the temperature
in MeV. See main text for details.}
\label{ssbar}
\end{figure}

In Fig.~\ref{ssbar}, we show the strange quark condensate normalized to its zero-temperature
value in different approximations as a function of the temperature.
The features are essentially the same as in Fig.~\ref{qqbar}, except 
that the electromagnetic effects are 
somewhat larger in this case. 

\begin{figure}[htb]
\centering
\includegraphics[width=0.48\textwidth]{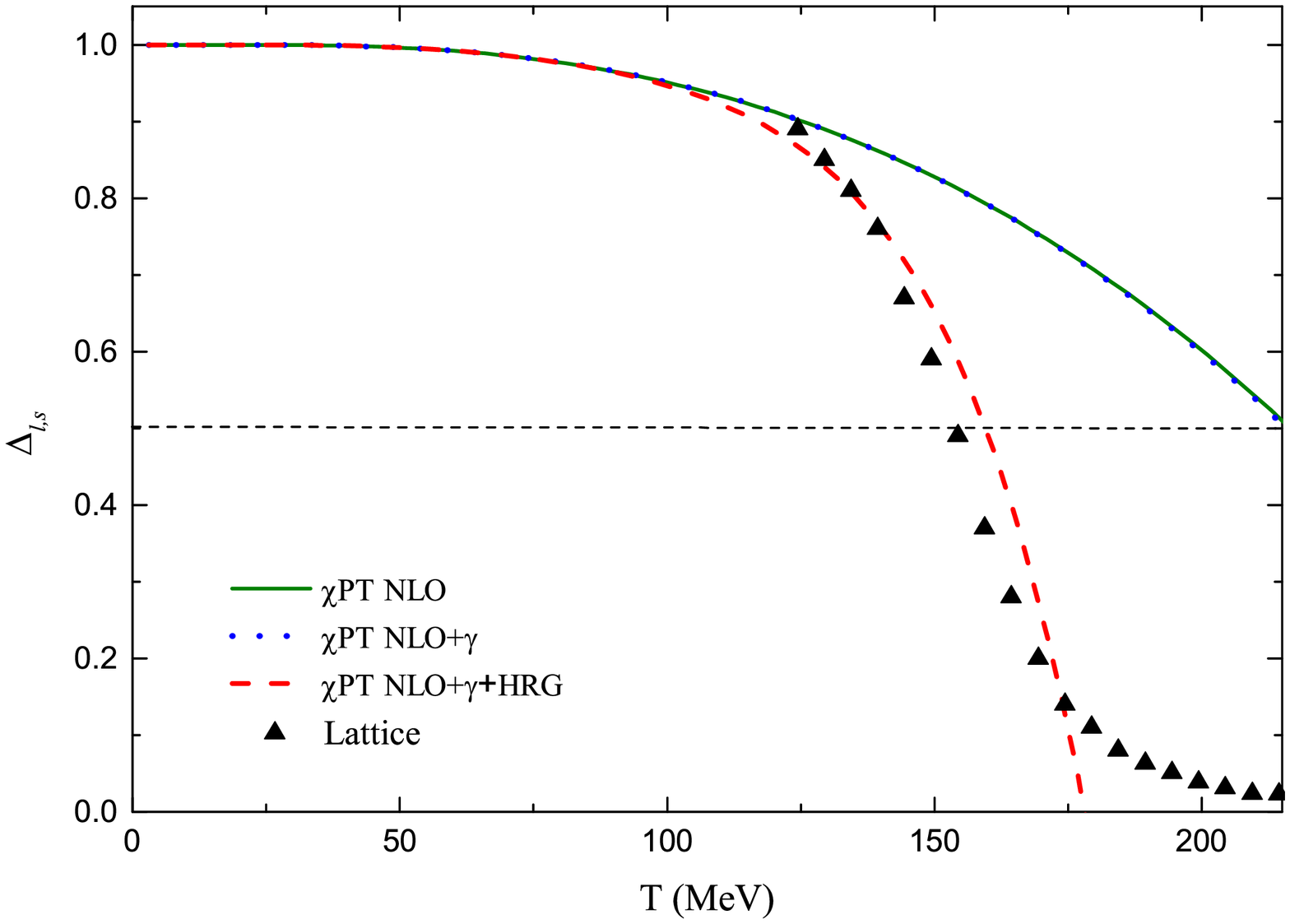}
\caption{$\Delta_{l,s}$ as a function of the temperature in MeV. 
The dotted horizontal curve indicates the value $\Delta_{l,s}={1\over2}$.}
See main text for details.
\label{lhs}
\end{figure}

In Fig.~\ref{lhs}, we plot the dimensionless 
quantity $\Delta_{l,s}$, which is defined as~\cite{tc3}
\bqa
\Delta_{l,s}&=&{\langle\bar{q}q\rangle_{T}-{m\over m_s}\langle\bar{s}s\rangle_{T}\over\langle\bar{q}q\rangle_{0}-{m\over m_s}\langle\bar{s}s\rangle_{0}}\;.
\eqa
The ratio of light quark mass $m$ and strange quark mass $m_s$ equals {1/30.21}, which is obtained
by using the bare values of $m_{\pi,0}$ and $m_{K,0}$. The green (blue) line is the $\mathcal{O}(p^4)$ results without (with) electromagnetic contributions. The red dotted line is for the results combined with the HRG model, which includes all the resonances states below 2.5 GeV in \cite{PDG}. For comparison, we also plot the lattice results in triangles from Ref.~\cite{tc3}. At low temperatures, the $\chi$PT predictions do converge well and the electromagnetic contributions are small. The agreement between the lattice results and the HRG is excellent all the way
up to $T\approx$ 170 MeV. For QCD with physical quark masses, there is no critical temperature.
However, one can define a crossover temperature in different ways. For example, the temperature at which the chiral condensate has decreased to half its vacuum value, or similarly
the temperature at which $\Delta_{l,s}$ has decreased to half its vacuum value.
It can also be defined as the temperature at which the quark susceptibilities has its peak.
Depending on the quantity, the crossover temperature in Ref.~\cite{tc3} is in the
150-170 MeV range. Using the definition $\Delta_{l,s}={1\over2}$ they obtain $T_{\rm pc}=157.3$ MeV.
Using the same definition, we obtain {160.1 MeV}, the dotted line in Fig.~\ref{lhs}
indiactes this value.
This crossover temperature is also very close to $T_{\rm pc}=161.2$ MeV obtained by a very recent HRG model analysis~\cite{peter}.
{Of course, one should bear in mind that the HRG model does not know about the
deconfined phase of QCD so the excellent agreement for $T_{\rm pc}$ obtained here
some extent be accidental. Likewise, the agreement with lattice data for temperatures above
approximately 150 MeV should be taken with a grain of salt for the same reason.}

\section*{Acknowledgements}
Q. Yu and H. Zhou have been supported by
the China Scholarship Council. They thank the Department of Physics at NTNU 
for kind hospitality during their stay.

\newpage

\pagebreak
\appendix

\section{Sum-integrals}
\label{sumint}
In the imaginary-time formalism, the four-momentum is $P=(p_0,{\bf p})$ with 
$P^2=p_0^2+{\bf p}^2$ and $p_0=2\pi nT$ being the Matsubara frequencies for bosons.
Loop integrals involve sums over Matsubara frequencies and 
integrals over spatial momenta. We use momentum-space dimensional regularization to regulate both infrared and ultraviolet divergences. The sum-integrals are defined as
\bqa
\sumint_P&=&T\sum_{p_0=2n\pi T}\int_p\;,
\eqa
where the sum is over Matsubara frequencies and
integrals over momenta are denoted by
\bqa
\int_p&=&\left({e^{\gamma_E}\Lambda^2\over4\pi}\right)^{\epsilon}\int{d^{d}p\over(2\pi)^{d}}\;, 
\eqa
where $d=3-2\epsilon$ and $\Lambda$ is an arbitrary momentum scale that coincides with the 
renormalization scale in the $\overline{\rm MS}$ scheme.
The one-loop integrals that appear in the calculations are of the form
\bqa
\label{defi0}
I_0^{\prime}(m^2)&=&-\sumint_P\log\left[P^2+m^2\right]\;,
\\ 
I_n(m^2)&=&\sumint_P{1\over(P^2+m^2)^n}\;,
\eqa
where the prime denotes differentiation with respect to the index $n$ evaluated at $n=0$.
They satisfy the relations
\bqa
{\partial\over\partial m^2}I_0^{\prime}&=&-I_1\;,
\hspace{1cm}
{\partial\over\partial m^2}I_n=-nI_{n+1}\;.
\label{reli}
\eqa
The sum-integral $I_n$ can be evaluated by standard contour-integration techniques.
We specifically need
\bqa\nonumber
I_0^{\prime}(m^2)&=&
{1\over2(4\pi)^2}\left({\Lambda\over m}\right)^{2\epsilon}\bigg[
\left({1\over\epsilon}+{3\over2}+{\cal O}(\epsilon)\right) m^4
 \\ &&+2J_0(\beta m)T^4\bigg]\;,
\label{i0p}
\\ \nonumber
I_1(m^2)&=&-{1\over(4\pi)^2}\left({\Lambda\over m}\right)^{2\epsilon}\bigg[
\left({1\over\epsilon}+1+{\pi^2+12\over12}\epsilon
\right)m^2
\\ && -J_1(\beta m)T^2\bigg]\;,
\label{i1}
\eqa
where the thermal integrals $J_n(\beta m)$ are defined as
\bqa\nonumber
J_n(\beta m)&=&{4e^{\gamma_{E}\epsilon}\Gamma({1\over2})\over\Gamma({5\over2}-n-\epsilon)}{m^{2\epsilon}\over T^{4-2n}}\int_0^{\infty}n(E_p){p^{4-2n-2\epsilon}\over E_p}dp\;,\\ &&
\label{defj}
\eqa
$n(E_p)={1/(e^{\beta E_p}-1)}$ is the Bose-Einstein (BE) distribution function, 
{$\beta$ is the inverse of temperature,} and $E_p=\sqrt{p^2+m^2}$.
The thermal integrals $J_n(x)$ satisfy the recursion relation
\bqa
xJ_n^{\prime}(x)&=&2\epsilon J_n(x)-2x^2J_{n+1}(x)\;.
\label{recj}
\eqa
For $\epsilon=0$ and in the limit $m\rightarrow0$, the thermal integrals behave as
\bqa
J_0&\rightarrow&{16\pi^4\over45}\;,
\\
\nonumber
J_1&\rightarrow&{4\pi^2\over3}-4\pi\beta m
-2\left(\log{\beta m\over4\pi}-{1\over2}+\gamma_E\right)(\beta m)^2\;,
\\ &&
\\ 
J_2&\rightarrow&
{2\pi\over\beta m}
+2\left(\log{\beta m\over4\pi}+\gamma_E\right)
\;.
\eqa
We notice that ${1\over2}I_0^{\prime}(0)$ reduces to ${\pi^2\over90}T^4$, which
is the Stefan-Boltzmann limit for the pressure of a massless bosonic degree of freedom.

\section{Evaluation of $I_{\rm sun}(m^2)$}
\label{suncalc}
In this appendix, we calculate 
\bqa
I_{\rm sun}(m^2)&\equiv&\sumint_{PQ}{1\over (P^2+m^2)(Q^2+m^2)(P+Q)^2}\;.
\label{sundeffie}
\eqa
We use the method of Bugrij and Shadura~\cite{bugrij}.
The sum-integrals over Euclidean momenta are replaced by integrals over four momenta 
$p=(p_0,{\bf p})$ in
Minkowski space, $\sumint_P\rightarrow -i\int_{M}$,
where integrals in Minkowski space are defined as 
\bqa
\int_{M}=\int_{-\infty}^{\infty}{dp_0\over2\pi}\int_p\;.
\eqa
The Euclidean propagator is replaced by the Minkowski propagator in the real-time formalism,
\bqa\nonumber
{1\over P^2+m^2}&\rightarrow&i\left({i\over p_0^2-E_p^2+i\epsilon}+n(|p_0|)2\pi\delta(p_0^2-E_p^2)\right)\;.
\\ &&
\eqa
The sum-integral $I_{\rm sun}(m^2)$ is then given by the
real part of the resulting expression. Some of the integrals involve one or more factors of 
the BE distribution.
The remaining integrals may conveniently be Wick-rotated back to 
Euclidean space, $\int_{M}\rightarrow i\int_P$, where the integral is defined as
\bqa
\int_P&=&
\left({e^{\gamma_E}\Lambda^2\over4\pi}\right)^{\epsilon}\int{d^{d+1}p\over(2\pi)^{d+1}}\;,
\eqa
with $d=3-2\epsilon$.
The term with zero thermal factors reads
\bqa\nonumber
{I}_{\rm sun}^{(0)}(m^2)&=&\int_{PQ}{1\over P^2(Q^2+m^2)[(P+Q)^2+m^2]}\\
&=&\int_P{1\over P^2}\Pi(P)\;,
\label{sundeff}
\eqa
where the superscript $(i)$ ($i=0,1,2,3$) 
of $I_{\rm sun}^{(i)}(m^2)$
denotes the number of BE factors and where
we have defined
\bqa
\Pi(P)&=&\int_Q{1\over(Q^2+m^2)[(P+Q)^2+m^2]}\;.
\label{bubble}
\eqa
\begin{widetext}
The bubble integrals can be calculated e.g. by using Feynman parameters,
\bqa
\Pi(P)&=&
{\pi\csc\epsilon\pi\over\Gamma(1-\epsilon)}
{(e^{\gamma_E}\Lambda^2)^{\epsilon}\over(4\pi)^2}\int_0^1dx\left[m^2+P^2x(1-x)\right]^{-\epsilon}\;.
\label{bubble2}
\eqa
Integrating first over four-momenta $P$ and then over $x$ yields
\bqa
{I}_{\rm sun}^{(0)}(m^2)&=&
{\pi\csc\epsilon\pi\over\Gamma(1-\epsilon)}
{(e^{\gamma_E}\Lambda^2)^{\epsilon}\over(4\pi)^2}
\int_0^1dx\int_P{1\over P^2}\left[m^2+P^2x(1-x)\right]^{-\epsilon}
=-{m^2\over(4\pi)^4}
\left({\Lambda\over m}\right)^{4\epsilon}e^{2\gamma_E\epsilon}
{(d-1)\pi^2\csc^2\epsilon\pi\over2(d-2)\Gamma^2(2-\epsilon)}\;.
\eqa
Expanding in powers of $\epsilon$ through order $\epsilon^0$ gives
\bqa
{I}_{\rm sun}^{(0)}(m^2)&=&-{m^2\over(4\pi)^4}\left({\Lambda\over m}\right)^{4\epsilon}
\left[{1\over\epsilon^2}+{3\over\epsilon}+7+{\pi^2\over6}+{\cal O}(\epsilon)\right]\;.
\label{t0sun}
\eqa
The terms with one thermal factor are
\bqa\nonumber
I_{\rm sun}^{(1)}(m^2)&=&
\int_{M}n(|p_0|)2\pi\delta({p_0^2-p^2})
\int_Q{1\over(Q^2+m^2)[(P+Q)^2+m^2]}\bigg|_{P^2=0}
+2\int_{M}n(|p_0|)2\pi\delta({p_0^2-E_p^2})
\\ &&\times
\int_Q{1\over(Q^2+m^2)(P+Q)^2}\bigg|_{P^2=-m^2}\:.
\label{t10}
\eqa
The integral over $Q$ in the first term is $\Pi(P)$ in Eq.~(\ref{bubble2}),
evaluated at $P^2=0$. The integral over $Q$ in the second term can be evaluated in the
same way. Expanding the resulting expressions in powers of $\epsilon$ yields
\bqa\nonumber
\int_Q{1\over(Q^2+m^2)[(P+Q)^2+m^2]}
&=&{1\over(4\pi)^2}\left({\Lambda\over m}\right)^{2\epsilon}
\left[{1\over\epsilon}-\int_0^1\log{m^2+x(1-x)P^2\over m^2}dx+{\cal O}(\epsilon)\right]\\ 
&=&{1\over(4\pi)^2}\left({\Lambda\over m}\right)^{2\epsilon}\left[{1\over\epsilon}
+{\cal O}(\epsilon)\right]\;,
\label{t11}
\\ \nonumber
\int_Q{1\over(Q^2+m^2)(P+Q)^2}
&=&{1\over(4\pi)^2}\left({\Lambda\over m}\right)^{2\epsilon}\left[{1\over\epsilon}-\int_0^1\log{xm^2+x(1-x)P^2\over m^2}dx
+{\cal O}(\epsilon)
\right]
\\ &=&
{1\over(4\pi)^2}\left({\Lambda\over m}\right)^{2\epsilon}\left[{1\over\epsilon}+2
+{\cal O}(\epsilon)\right]\;.
\label{t12}
\eqa
Substituting Eqs.~(\ref{t11}) and~(\ref{t12}) into Eq.~(\ref{t10}) 
and integrating over $p_0$ yields
\bqa
{I}_{\rm sun}^{(1)}(m^2)&=&
{1\over(4\pi)^2}\left({\Lambda\over m}\right)^{2\epsilon}
\int_p\left[{n(p)\over p}{1\over\epsilon}+{2n(E_p)\over E_p}\left({1\over\epsilon}+2\right)
\right]\;.
\label{1thermal}
\eqa
The terms with two thermal factors are
\bqa\nonumber
{I}_{\rm sun}^{(2)}(m^2)&=&
\int_{M}n(|p_0|)2\pi\delta({p_0^2-E_p^2})
\int_{M}
n(|q_0|)2\pi\delta(q_0^2-E_q^2)
{1\over(p_0+q_0)^2-({\bf p}{+}{\bf q})^2}
\\ &&
+2\int_{M}n(|p_0|)
2\pi\delta(p_0^2-p^2)
\int_{M}n(|q_0|)2\pi
\delta(q_0^2-E_q^2)
{1\over(p_0+q_0)^2-({\bf p}{+}{\bf q})^2-m^2}\;.
\eqa
This integral is convergent in three dimensions so we set $\epsilon=0$.
We first integrate over $p_0$ and $q_0$, which yields
\bqa\nonumber
I_{\rm sun}^{(2)}(m^2)&=&
{1\over2}
\int_{pq}{n(E_p)n(E_q)\over E_pE_q}
\left[{1\over(E_p+E_q)^2-({\bf p}+{\bf q})^2}+{1\over(E_p-E_q)^2-({\bf p}{+}{\bf q})^2}
\right]\\
&&
+\int_{pq}{n(p)n(E_q)\over pE_q}
\left[{1\over({p}+E_q)^2-({\bf p}+{\bf q})^2-m^2}
+{1\over({p}-E_q)^2-({\bf p}{+}{\bf q})^2-m^2}
\right]\;.
\label{second}
\eqa
Averaging over the angle between ${\bf p}$ and ${\bf q}$ gives
\bqa
I_{\rm sun}^{(2)}(m^2)&=&{8\over(4\pi)^4}\int_0^{\infty}{pn(E_p)qn(E_q)\over E_pE_q}
\log\bigg|{(p-q)^2\over(p+q)^2}\bigg| dpdq\;,
\label{2thermal}
\eqa
where we notice that the angular average of the term in the second line of Eq.~(\ref{second}) 
vanishes. Finally, 
the term with three thermal factors, $I_{\rm sun}^{(3)}(m^2)$, is purely imaginary and is dropped.
Adding Eqs.~(\ref{t0sun}), ~(\ref{1thermal}), and ~(\ref{2thermal}), we obtain
the result for the setting-sun diagram
\bqa\nonumber
I_{\rm sun}(m^2)&=&-
{m^2\over(4\pi)^4}\left({\Lambda\over m}\right)^{4\epsilon}
\left[{1\over\epsilon^2}+{3\over\epsilon}+7+{\pi^2\over6}\right]+
{1\over(4\pi)^2}\left({\Lambda\over m}\right)^{2\epsilon}
\int_p\left[{n(p)\over p}{1\over\epsilon}
+{2n(E_p)\over E_p}\left({1\over\epsilon}+2\right)\right]
\\&&
+{8\over(4\pi)^4}\int_0^{\infty}{pn(E_p)qn(E_q)\over E_pE_q}
\log\bigg|{(p-q)^2\over(p+q)^2}\bigg| dpdq\;.
\label{setsundef}
\eqa
\end{widetext}

\section{One-loop vacuum energy}
\label{vacuum}
In this appendix, we calculate the vacuum energy $V$ including electromagnetic effects.
From $V$ it is easy to derive the one-loop corrections to the quark condensates
that are needed in our finite-temperature formulas.
The ${\cal O}(p^2)$ contribution is
\bqa
V_{0}&=&-{1\over2}f^2(m_{\pi,0}^2+2m_{K,0}^2)\;.
\eqa
\begin{widetext}
The ${\cal O}(p^4)$ contribution from the loops is
\bqa
\nonumber
V_1&=&{1\over2}\int_P\log[P^2+m_{\pi,0}^2]+
\int_P\log[P^2+m_{\pi^{\pm},0}^2]
+\int_P\log[P^2+m_{K^{\pm},0}^2]
+\int_P\log[P^2+m_{K,0}^2]
\\ && 
+{1\over2}\int_P\log[P^2+m_{\eta,0}^2]
+(d-1)
\int_P\log[P^2]
\;,
\eqa
where the last term comes from the photons and ghost and vanishes at zero temperature.
The ${\cal O}(p^4)$ counterterm contribution is
\bqa\nonumber
V_1^{\rm ct}&=&
-\left(4L_6-2L_8-H_2\right)
\left(m_{\pi,0}^2+2m_{K,0}^2\right)^2
-4(2L_8+H_2)\left(m_{\pi,0}^4+2m_{K,0}^4\right)
\\ \nonumber
&&-{4e^2f^2m_{\pi,0}^2\over3}\left[K_7+K_{8}+{4\over3}K_{9}+{4\over3}K_{10}\right]
-{8e^2f^2m_{K,0}^2\over3}\left[K_7+K_{8}+{1\over3}K_{9}+{1\over3}K_{10}\right]
\\ &&
-{4e^4f^4\over9}\left[K_{15}+K_{16}+K_{17}\right]\;.
\eqa
After renormalization, we find the vacuum energy for three-flavor $\chi$PT to
${\cal O}(p^4)$ including electromagnetic effects.
\bqa
\nonumber
V_{0+1}&=&-{1\over2}f^2(m_{\pi,0}^2+2m_{K,0}^2)
-\left(4L_6^r-2L_8^r-H_2^r\right)
\left(m_{\pi,0}^2+2m_{K,0}^2\right)^2
\\ && \nonumber
-4(2L_8^r+H_2^r)\left(m_{\pi,0}^4+2m_{K,0}^4
\right)
-{{m_{\pi,0}^4}\over4(4\pi)^2}\left(\log{\Lambda^2\over m_{\pi,0}^2}+{1\over2}\right)
-{{m_{\pi^{\pm},0}^4}\over2(4\pi)^2}
\left(\log{\Lambda^2\over m_{\pi^{\pm},0}^2}+{1\over2}\right)
\\ && \nonumber
-{{m_{K^{\pm},0}^4}\over2(4\pi)^2}\left(\log{\Lambda^2\over m_{K^{\pm},0}^2}+{1\over2}\right)
-{{m_{K,0}^4}\over2(4\pi)^2}\left(\log{\Lambda^2\over m_{K,0}^2}+{1\over2}\right)
-{{m_{\eta,0}^4}\over4(4\pi)^2}\left(\log{\Lambda^2\over m_{\eta,0}^2}+{1\over2}\right)
\\ && \nonumber
-{4e^2f^2m_{\pi,0}^2\over3}\left[K_7^r+K_{8}^r+{4\over3}K_{9}^r+{4\over3}K_{10}^r
\right]
-{8e^2f^2m_{K,0}^2\over3}\left[K_7^r+K_{8}^r+{1\over3}K_{9}^r+{1\over3}K_{10}^r
\right]
\\
&&-{4e^4f^4\over9}\left[K_{15}^r+K_{16}^r+K_{17}^r
\right]
\;.
\eqa
The light and $s$-quark condensates to ${\cal O}(p^4)$
in the vacuum are
\bqa
\langle\bar{q}q\rangle_0&=&
-2f^2B_0\left[1
+{m^2_{\pi,0}\over f^2}\left(16L^r_6+8L^r_8+4H^r_2
+{1\over2(4\pi)^2}\log{\Lambda^2\over m^2_{\pi,0}}\right)
+{m^2_{K,0}\over f^2}\left(32L^r_6+
{1\over2(4\pi)^2}\log{\Lambda^2\over m^2_{K,0}}\right)
\right. \nonumber\\&&\left. \nonumber
+{m^2_{\pi^\pm,0}\over(4\pi)^2f^2}\log{\Lambda^2\over m^2_{\pi^\pm,0}}
+{m^2_{K^\pm,0}\over2(4\pi)^2f^2}\log{\Lambda^2\over m^2_{K^\pm,0}}
+{m^2_{\eta,0}\over6(4\pi)^2f^2}\log{\Lambda^2\over m^2_{\eta,0}}
\right]
\\ &&
\label{light1}
-{16e^2f^2B_0\over3}\left(K_7^r+K_8^r+{5\over6}K_9^r+{5\over6}K_{10}^r\right)\;, \\
\langle\bar{s}s\rangle_0&=&
-f^2B_0\left[1
+{4m^2_{\pi,0}\over f^2}\left(4L^r_6-2L^r_8-H^r_2\right)
+{m^2_{K,0}\over f^2}\left(32L^r_6+16L^r_8+8H^r_2+{1\over(4\pi)^2}\log{\Lambda^2\over m^2_{K,0}}\right)
\right. \nonumber\\&&\left.
+{m^2_{K^\pm,0}\over(4\pi)^2f^2}\log{\Lambda^2\over m^2_{K^\pm,0}}
+{2m^2_{\eta,0}\over3(4\pi)^2f^2}\log{\Lambda^2\over m^2_{\eta,0}}
\right]
-{8e^2f^2B_0\over3}\left(K_7^r+K_8^r+{1\over3}K_9^r+{1\over3}K_{10}^r\right)
\;.
\label{slast}
\eqa

\end{widetext}
Note that quark condensates depend on the coupling $H_2^r$, which is unphysical in the
sense that it arises from a contact term in the ${\cal O}(p^4)$ Lagrangian.

\section{Meson masses and pion-decay constant}
\label{mesonmass}

In this appendix, we list the meson masses to one-loop order including the leading electromagnetic
effects, i.e. through $e^2$. The meson masses without electromagnetic corrections were calculated
in Ref.~\cite{gasser2}, while electromagnetism was included in Ref.~\cite{urech1}.
The meson masses without electromagnetic corrections are
\begin{widetext}
\bqa \nonumber
{M}^2_{\pi}&=&
m^{2}_{\pi,0}\left[1
-{m^2_{\pi,0}\over f^2}\left(8L^r_4+8L^r_5-16L^r_6-16L^r_8
+{1\over2(4\pi)^2}\log\frac{\Lambda^2}{m^2_{\pi,0}}\right)
-{m^2_{K,0}\over f^2} \left(16L^r_4-32L^r_6 \right)
\right. \\ && \left.
+ {m^2_{\eta,0}\over6(4\pi)^2f^2}\log{\Lambda^2\over m^2_{\eta,0}} \right]\;,
\label{pi00}
\\ 
{M}^2_{K}&=&
m^2_{K,0}\left[1
-{m^2_{\pi,0}\over f^2}(8L^r_4-16L^r_6)
-{m^2_{K,0}\over f^2}(16L^r_4+8L^r_5-32L^r_6-16L^r_8)
-\frac{m^2_{\eta,0}}{3(4\pi)^2f^2}\log\frac{\Lambda^2}{m^2_{\eta,0}}\right]\;,
\label{k00}
\\ 
{M}^2_{\eta}&=&
m^2_{\eta,0}\left[
1
-{m^2_{\pi,0}\over f^2}\left(8L^r_4-16L^r_6
+{1\over6(4\pi)^2}\log\frac{\Lambda^2}{m^2_{\eta,0}}\right)
-{m^2_{K,0}\over f^2}\left(16L^r_4-32L^r_6+
{1\over(4\pi)^2}\log\frac{\Lambda^2}{m^2_{K,0}}\right)
 \right. \nonumber\\ && \left.
-{m^2_{\eta,0}\over f^2}\left(8L^r_5-{2\over3(4\pi)^2}\log\frac{\Lambda^2}{m^2_{\eta,0}}
\right)\right]\;
+L^r_7\frac{128(m^2_{\pi,0}-m^2_{K,0})^2}{3f^2}
\nonumber\\&&
+L^r_8\frac{16}{3f^2}(3m^4_{\pi,0}-8m^2_{\pi,0}m^2_{K,0}+8m^4_{K,0})
+\frac{m^4_{\pi,0}}{2(4\pi)^2f^2}\log\frac{\Lambda^2}{m^2_{\pi,0}}
-\frac{m^2_{\pi,0}m^2_{K,0}}{3(4\pi)^2f^2}\log\frac{\Lambda^2}{m^2_{K,0}}.
\eqa
{After including the one-loop $\chi$PT contribution to meson masses and}
electromagnetic effects up to order $e^2$, the {charged and neutral} meson masses are
\bqa \nonumber
m^2_{\pi^0}&=&{M^{2}_{\pi}}+e^{2}m^{2}_{\pi,0}\left(
-\frac{8}{3}K^r_{1}-\frac{8}{3}K^r_{2}+4K^r_3-2K^r_4-\frac{20}{9}K^r_{5}\nonumber 
-\frac{20}{9}K^r_{6}+\frac{8}{3}K^r_{7}+\frac{8}{3}K^r_{8}+\frac{20}{9}K^r_{9}+\frac{20}{9}K^r_{10}
\right)\nonumber
\\&&
+2\frac{Ce^2m^2_{\pi,0}}{(4\pi)^2f^4}
\left(1-\log\frac{\Lambda^2}{m^2_{\pi,0}}\right)\;,
\label{mes1}
\\ \nonumber
m^2_{\pi^\pm}&=&{M^{2}_{\pi}}+e^{2}m^{2}_{\pi,0}\left(
-\frac{8}{3}K^r_{1}-\frac{8}{3}K^r_{2}-\frac{20}{9}K^r_{5}-\frac{20}{9}K^r_{6}+\frac{8}{3}K^r_{7}\nonumber 
+\frac{20}{3}K^r_{8}+\frac{20}{9}K^r_{9}+\frac{92}{9}K^r_{10}+8K^r_{11}\right)
\\ &&+8e^2m^2_{K,0} K^r_8+{e^2m^2_{\pi,0}\over(4\pi)^2}\left(4+3\log {\Lambda^2\over m^2_{\pi,0}}\right)
- L^r_4\frac{16Ce^2}{f^4}(m^2_{\pi,0}+2m^2_{K,0})-L^r_5 \frac{16Ce^2m^2_{\pi,0}}{f^4}\nonumber\\&&
{+4\frac{Ce^2m^2_{\pi,0}}{(4\pi)^2f^4}\log\frac{\Lambda^2}{m^2_{\pi,0}}}
+2\frac{Ce^2m^2_{K,0}}{(4\pi)^2f^4}\log\frac{\Lambda^2}{m^2_{K,0}}\;,
\eqa
\bqa
\nonumber
m^2_{K^{\pm}}&=&{M^2_{K}}+e^{2}m^{2}_{K,0}\left(
-\frac{8}{3}K^r_{1}-\frac{8}{3}K^r_{2}-\frac{20}{9}K^r_{5}-\frac{20}{9}K^r_{6}+\frac{8}{3}K^r_{7}+\frac{32}{3}K^r_{8}
+\frac{8}{9}K^r_{9}+\frac{80}{9}K^r_{10}+8K^r_{11}\right)
\\ \nonumber&&
+e^2m^2_{\pi,0}\left( 4K^r_8+\frac{4}{3} K^r_9+\frac{4}{3} K^r_{10}\right)
+{e^2m^2_{K,0}\over(4\pi)^2}\left(4+3\log {\Lambda^2\over m^2_{K,0}}\right)
- L^r_4\frac{16Ce^2}{f^4}(m^2_{\pi,0}+2m^2_{K,0})
\nonumber\\&&
-L^r_5 \frac{16Ce^2m^2_{K,0}}{f^4}
+2\frac{Ce^2m^2_{\pi,0}}{(4\pi)^2f^4}\log\frac{\Lambda^2}{m^2_{\pi,0}}
{+4\frac{Ce^2m^2_{K,0}}{(4\pi)^2f^4}\log\frac{\Lambda^2}{m^2_{K,0}}},
\\
m^2_{K^0}&=&{M^2_{K}}
+\frac{8}{9}e^{2}m^2_{K,0}\left(-3K^r_{1}-3K^r_{2}-K^r_{5}-K^r_{6}
+3K^r_{7}+3K^r_{8}+K^r_{9}+K^r_{10}\right)\;,
\eqa
\bqa\nonumber
m^2_{\eta}&=& {M^2_{\eta}}+e^2m^2_{\eta,0}\left(-\frac{8}{3}K^r_{1}-\frac{8}{3}K^r_{2}+\frac{4}{3}K^r_3-\frac{2}{3}K^r_4-\frac{4}{3}K^r_{5}\nonumber
-\frac{4}{3}K^r_{6}+\frac{8}{3}K^r_{7}+\frac{8}{3}K^r_{8}
+\frac{8}{9}K^r_{9}+\frac{8}{9}K^r_{10}\right)\nonumber\\&&
+\frac{4}{9}e^2 m^{2}_{\pi,0}( K^r_{9}+K^r_{10})
-\frac{2}{3}\frac{Ce^2m^2_{\pi,0}}{(4\pi)^2f^4}
\left(1-\log\frac{\Lambda^2}{m^2_{\pi,0}}\right)
+\frac{Ce^2m^2_{K,0}}{3(4\pi)^2f^4}\left(1-4\log\frac{\Lambda^2}{m^2_{K,0}}
\right)\;.
\label{mes5}
\eqa
Finally, including electromagnetic effects, the {neutral} pion-decay constant $f_{\pi^0}$ is
\bqa
f_{\pi^0}
&=&f\left[1  +\left(4{L}_4^r
+4L_5^r\right){m_{\pi,0}^2\over f^2}+8L_4^r{m_{K,0}^2\over f^2}
+e^2\left({4\over3}K^r_1+{4\over3}K^r_2-2K^r_3+K^r_4+{10\over9}K^r_5+{10\over9}K^r_6\right)
\right. \nonumber\\&&\left.
+{m_{\pi^\pm,0}^2\over(4\pi)^2f^2}\log{\Lambda^2\over m_{\pi^\pm,0}^2}
+{m_{K^\pm,0}^2\over4(4\pi)^2f^2}\log{\Lambda^2\over m_{K^\pm,0}^2}
+{m_{K,0}^2\over4(4\pi)^2f^2}\log{\Lambda^2\over m_{K,0}^2}
\right]\;.
\label{fpi}
\eqa
\end{widetext}

\bibliographystyle{apsrmp4-1}

\begin{thebibliography}{}


\bibitem{wein}
  S. Weinberg, 
  Physica A {\bf 96}, 327 (1979).

\bibitem{gasser1}
J. Gasser and H. Leutwyler, Ann. Phys. {\bf 158}, (142) (1984).

\bibitem{gasser2}
J. Gasser and H. Leutwyler, Nucl. Phys. B {\bf 250}, 465 (1985).






\bibitem{fearing}
H. W. Fearing and S. Scherer,
Phys. Rev. D {\bf 53}, 315 (1996).


\bibitem{bein}
  J. Bijnens, G. Colangelo and G. Ecker,
  Ann. Phys. {\bf 280}, 100 (2000).

\bibitem{bein2}
 J. Bijnens, G. Colangelo and G. Ecker,
JHEP {\bf 02} 020 (1999).


\bibitem{bijnensreview}
J. Bijnens and G. Ecker,
Ann. Rev. Nucl. Part. Sci. {\bf 64}, 149 (2014).



\bibitem{e2chi}
G. Ecker, J. Gasser, A. Pich, and E. de Rafael,
Nucl. Phys. B {\bf 321}, 311 (1989).

\bibitem{urech1}     
R. Urech, Nucl. Phys. B {\bf 433}, 234 (1995).


\bibitem{meis1}

U.-G. Meissner, G. M\"uller and S. Steininger, 
Phys. Lett. B {\bf 406} (1997). 

\bibitem{meis2}
U.-G. Meissner, G. M\"uller and S. Steininger,
Phys. Lett. B {\bf 407} 454 (1997) (erratum).



\bibitem{urech2}
M. Knecht and R. Urech,
Nucl. Phys. B {\bf 519}, 329 (1998).




\bibitem{gasser3}
J. Gasser and H. Leutwyler,
Phys. Lett. B {\bf 184}, 83 (1987).

\bibitem{gas4}
J. Gasser and H. Leutwyler,
Phys. Lett. B {\bf 188}, 477 (1987).

\bibitem{finitet} 
P. Gerber and H. Leutwyler,
Nucl. Phys. B {\bf 321}, 387 (1989).





\bibitem{tc1}
Y. Aoki, Z. Fodor, S. Katz, and K. Szabo, 
Phys. Lett. B {\bf 643}, 46 (2006).

\bibitem{tc2}
Y. Aoki, S. Borsanyi, S. D\"urr, Z. Fodor, S.D. Katz et al., 
JHEP {\bf 09} 06, 088 (2009).

\bibitem{tc3}
S. Borsanyi et al. (Wuppertal-Budapest Collaboration),
JHEP {\bf 10} 09, 073 (2010).


\bibitem{tc4}
A. Bazavov, T. Bhattacharya, M. Cheng, C. DeTar, H. Ding et al., 
Phys. Rev. D {\bf 85}, 054503 (2012).

\bibitem{hrg1}
F. Karsch, K. Redlich and A. Tawfik, Eur. Phys. J. C 29, 549 (2003).


\bibitem{hrg2}
F. Karsch, K. Redlich and A. Tawfik, Phys. Lett. B 571, 67 (2003).

\bibitem{bla}
J. Jankowski, D. Blaschke, and M. Spaliński
Phys. Rev. D {\bf 87} 105018 (2013).

\bibitem{peter}
D. Biswas, P. Petreczky, and S. Sharma,
e-Print: 2206.04579.


\bibitem{nicola}
A. Nicola and R. Torres Andrés
Phys. Rev. D {\bf 83}, 076005 (2011).

\bibitem{massnicola}
A. Nicola and R. Torres Andrés
Phys. Rev. D {\bf 89}, 116009 (2014).

\bibitem{bpdamp}
E. Braaten and R. D. Pisarski,
Phys. Rev. D {\bf 42}, 2156 (1990).


\bibitem{softampl}
E. Braaten and R. D. Pisarski,
Nucl. Phys. B {\bf 337}, 569 (1990).





\bibitem{gil1}
E. Braaten and R. D. Pisarski, 
Phys. Rev. D {\bf 45}, R1827 (1992).

\bibitem{gil2}
J. Frenkel and  J. C. Taylor,
Nucl. Phys. B {\bf 374}, 156 (1992).



\bibitem{ericallt}
E. Braaten,
Can. J. Phys. {\bf 71}, 215 (1993) 215.









 








\bibitem{gas}
J. Gasser, A. Rusetsky, and I. Scimemi,
Eur. J. Phys. C {\bf 32}, 97 (2003). 
        



\bibitem{MartinCamalich:2010fp}
J.~Martin Camalich, L.~S.~Geng and M.~J.~Vicente Vacas,
Phys. Rev. D {\bf82}, 074504 (2010).
  








\bibitem{PDG}
P. A.~Zyla \textit{et al.} [Particle Data Group],
PTEP \textbf{2020}, no.8, 083C01 (2020).







\bibitem{jamin}
M. Jamin, 
Phys. Lett. B {\bf 538}, 71 (2002).

\bibitem{Gorghetto:2018ocs}
M.Gorghetto and G.Villadoro,
JHEP \textbf{03}, 033 (2019).

\bibitem{Bijnens:1996kk}
J.~Bijnens and J.~Prades,
Nucl. Phys. B \textbf{490}, (1997).


\bibitem{endrodihrg}
G. Endrodi, JHEP {\bf 04}, 23 (2013).





















\bibitem{bugrij}
A. I. Bugrij and V. N. Shadura, hep-th/9510232.

\end{thebibliography}

\end{document}